\documentclass[12pt]{article}

\usepackage{amsmath}
\usepackage{indentfirst}
\usepackage{amsfonts,amssymb}
\usepackage{mathrsfs}
\usepackage{CJK,CJKnumb}
\usepackage{appendix}
\usepackage{amsthm}
\usepackage{color}
\usepackage{float}
\usepackage{booktabs}
\usepackage{multirow}
\usepackage{mathrsfs}
\usepackage{graphicx}
\usepackage{subfigure}
\usepackage{subfig}
\usepackage{caption}
\makeatletter
\newcommand{\rmnum}[1]{\romannumeral #1} \newcommand{\Rmnum}[1]{\expandafter\@slowromancap\romannumeral #1@}
\makeatother
\usepackage{chngcntr}
\counterwithin{figure}{section}

\newtheorem{Definition}{Definition}[section]
\newtheorem{Theorem}{Theorem}[section]
\newtheorem{Lemma}{Lemma}[section]

\numberwithin{equation}{section}

\begin{document}

\title{
        { Optimal investment, consumption and life insurance decisions for households with consumption habits under the health shock risk }
            \footnotetext{Email address: zzhao\underline{ }math@stu.xju.edu.cn (Z. Zhao) ; liuwei.math@xju.edu.cn (W. Liu) ;\\  xytang\underline{ }math@stu.xju.edu.cn (X. Tang).}
            }

\author{Zhen Zhao , \quad Wei Liu$^*$ , \quad Xiaoyi Tang \\
  \\
 College of Mathematics and System Sciences\\
      Xinjang University\\
      Urumqi, Xinjiang 830046\\
      People's Republic of China \\
     }

\date{}

\maketitle

\def\sA{{\mathscr A}}
\def\sB{{\mathscr B}}
\def\sC{{\mathscr C}}
\def\sD{{\mathscr D}}
\def\sF{{\mathscr F}}
\def\sG{{\mathscr G}}
\def\sH{{\mathscr H}}
\def\sM{{\mathscr M}}
\def\sP{{\mathscr P}}
\def\sQ{{\mathscr Q}}
\def\sR{{\mathscr R}}
\def\sS{{\mathscr S}}
\def\sT{{\mathscr T}}
\def\sW{{\mathscr W}}
\def\sX{{\mathscr X}}
\def\sY{{\mathscr Y}}

\vspace{0.6cm}

\begin{abstract}
\quad This paper investigates the optimal investment, consumption, and life insurance strategies for households under the impact of health shock risk. Considering the uncertainty of the future health status of family members, a non-homogeneous Markov process is used to model the health status of the breadwinner. Drawing upon the theory of habit formation, we investigate the influence of different consumption habits on households' investment, consumption, and life insurance strategies. Based on whether the breadwinner is alive or not, we formulate and solve the corresponding Hamilton-Jacobi-Bellman (HJB) equations for the two scenarios of breadwinner survival and breadwinner's demise, respectively, and obtain explicit expressions for the optimal investment, consumption, and life insurance strategies. Through sensitivity analysis, it has been shown that the presence of health shocks within households has a negative impact on investment and consumption decisions, while the formation of consumption habits increases household propensity for precautionary savings.
\end{abstract}

\vspace{0.5cm}


\vspace{0.5cm}


\newpage

\section{Introduction}

In the study of asset allocation within households, the health shock risk faced by family members is a crucial factor. Firstly, the presence of health shocks can potentially necessitate medical treatments and care for family members, thereby leading to an escalation in healthcare expenditure for the household. According to Wang et al. (2023), an unexpected health shock increases household medical expenditure by \$ 2,647 over five years. Secondly, health shocks can prevent family members from working normally or needing time off for treatment and recovery. As a result, household income is reduced or interrupted, impacting household finances. Lenhart (2019) confirmed that annual household labor income fell $\pounds$1,181.40 after health shocks. Households struggle to maintain daily expenses and repayment commitments as incomes fall. Moreover, households may need to borrow to pay for medical or other health-related expenses. This will increase the debt burden of households, perhaps leading to excessive interest payments and financial distress in the long-term. Therefore, health shock risk can cause economic instability. It reduces the consumption power of the household to meet basic needs or desired life goals. Exploring the households' investment and consumption under the influence of health shock risk is of great practical benefit. Useful for guiding the household in risk management and financial planning, maintaining the households financial stability.

\vspace{0.3cm}

Health shock risk refers to the possibility that an individual's health status may be impaired due to various factors such as illness, accidental injury, genetic problems, etc. This may lead households to be more conservative in their investments. Edwards (2008) indicated that when individuals sense health risks, they tend to reduce the proportion of risky investments. Based on the analysis of socioeconomic data in Germany, Decker and Schmitz (2016) found that health shocks increase the aversion to risk within households, subsequently affecting their investment decisions. Other studies, such as Riphahn (1999), Berkowitz and Qiu (2006), Halla and Zweim¨¹ller (2013), Lenhart (2019), and Rice and Robone (2022), have also extensively explored the relationship between health shocks and household investment behavior. Drawing upon this foundation, Hambel et al. (2017) discussed the issue of household investment, consumption, and insurance under permanent health shocks, with the optimization goal of maximizing the expected utility of consumption and terminal wealth. Hambel (2020) employs the Cobb-Douglas utility function to analyze the optimal investment behavior of agents with stochastic wages under health shock risk. He discovered that the relative reduction in optimal investment depends on the age at which agents experience a health shock. Barucci et al. (2023) concentrated on the impact of health shocks on the labor market. Kraft and Weiss (2023) examined the problem of portfolio selection under health shocks using COVID-19 as an example. Using a dynamic programming approach, they confirmed that health shocks lead agents to significantly reduce the size of their stock investments.

\vspace{0.3cm}

Traditional consumption theory generally assumes that people's consumption depends on their current income and wealth. However, according to Duesenberry (1949), people's current consumption is influenced not only by their income levels but also by their previous consumption patterns. Consumption habits can be understood as preferences and habitual tendencies towards certain levels of consumption. For example, when individuals experience an increase in income, they may tend to maintain their previous level of consumption rather than immediately increase spending. Similarly, when individuals encounter a reduction in income, they may maintain their previous level of consumption rather than immediately cut back. This may explain why individuals experience smoother variation in consumption than in income. Moreover, individuals' consumption patterns have a direct impact on their savings and investment tendencies, thereby influencing their financial standing and future purchasing capacity.
Hicks (1965), Pollak (1970), and Ryder and Heal (1973) modeled consumption habits in the von Neumann-Morgenstern setting by introducing a form of temporal inseparability. They modeled consumption habits as weighted indicators of past consumption. Sundaresan (1989) applied consumption habits theory to investment-consumption problems and explained the consumption smoothing phenomenon. Detemple and Zapatero (1992) extended the model to the case of stochastic coefficients. They were the first to consider the impact of consumption habits on individual optimal investment and consumption. Polkovnichenko (2006) investigated the impact of additive consumption habits on portfolio choice and suggested that youthful households should adopt a more conservative investment strategy. Using dynamic programming, Liu et al. (2020) proved that the presence of consumption habits reduces the assets available for allocation by introducing consumption habits into an individual investment, consumption, and life insurance decision model. Boyle et al. (2022) evaluated the impact of consumption habits on the demand for whole-life insurance in a single-earner household. They showed that consumption habits can change the motivation of bequests. Further evidence of the impact of consumption habits on investment decisions can be found in the literature by Ben-Arab et al. (1996), Pirvu and Zhang (2012), and Kraft et al. (2017). The above studies show that the impact of consumption habits on individual and household investment, consumption and insurance decisions cannot be ignored.

\vspace{0.3cm}

If there are dependents in the households, the breadwinners' death means an interruption in the households' income. Life insurance can help households cope with such economic shocks and provide some financial security for dependents. The seminal work of Yaari (1965) and Richard (1975) studied the problem of optimizing investment, consumption, and life insurance when the life distribution is known. Based on these findings, much research has been carried out to explore the impact of life insurance on the economic stability and future security of individuals and households. Charupat and Milevsky (2002), Moore and Young (2006), Nielsen and Steffensen (2008), Pirvu and Zhang (2012), Mousa et al. (2016), Ye (2019), Peng and Li (2023) and other scholars discussed the optimal strategy of life insurance from the perspective of individuals. The survey by Kwak et al. (2011) focused on a single breadwinner household and used the martingale method to obtain optimal investment, consumption, and life insurance strategies. Wang et al. (2021) further considered the income risk of a household with only one breadwinner and one dependent. Wei et al. (2020) investigated the impact of longevity interdependence on couples life insurance. Moreover, Luciano et al. (2008), Bruhn and Steffensen (2011), Bayraktar and Young (2013), Liang and Zhao (2016), Lee and Cha (2018), and other researchers also made relevant studies. They highlighted the crucial influence of life insurance on household economic and investment decisions.

\vspace{0.3cm}

In this paper, we study the impact of health shock risk on the optimal investment, consumption, and life insurance strategies of households in terms of income and mortality. By considering a household model with one breadwinner and one dependent, a non-homogeneous Markov process is used to model the health status of the breadwinner. The stochastic transformation of health status represents the health shock risk faced by the household. Taking into account the influence of past consumption behavior on current optimal investment, consumption, and life insurance strategies, the study models the consumption habits of the household using exponentially weighted past consumption rates, examining the differences in asset allocation strategies under different consumption habits. Based on the optimal criterion of maximum expected utility of cumulative consumption and terminal wealth of the household, two situations are divided according to the survival state of the breadwinner, and the optimal control model is established, respectively.
Explicit expressions for optimal investment, consumption, and life insurance strategies are derived by employing dynamic programming methods.
By sensitivity analysis of important parameters, we reveal that in the presence of health shocks, households tend to adopt conservative investment strategies, spend less on consumption and more on life insurance. Additionally, the formation of consumption habits allows households to initially reduce their consumption expenditures, while increasing their savings to meet future high consumption needs.

\vspace{0.3cm}

The contributions of this paper are as follows: (\rmnum{1}) Reflect health shocks through non-homogeneous Markov processes and consider the influence of health status of the breadwinner in the framework of household asset allocation. Optimal strategies are derived by formulating and solving control models separately in two scenarios based on the survival status of the breadwinner. (\rmnum{2}) By replacing traditional consumption functions with consumption habits, this papaer explores the impact of consumption habits on optimal household investment, consumption, and life insurance decisions. (\rmnum{3}) The critical illness model is taken as an example to verify the research conclusion. Some economic insights are gained through numerical examples. For example, health shock risk has a negative impact on investment and consumption decisions.

\vspace{0.3cm}

The rest of this paper is structured as follows. In Section 2, the health status, insurance markets, consumption habits, and stochastic control problems are characterized. In Section 3, the models are built and the explicit solutions for the optimisation problems are obtained. In Section 4, the critical illness model is used as an example and the parameters are calibrated. In Section 5, a sensitivity analysis of the essential parameters is performed. In Section 6, the full text is summarized. The detailed solving procedure is given in the Appendix.

\section{Model setup and problem formulation}

\renewcommand{\theequation}{\thesection.\arabic{equation}}

This paper delves into the investment, consumption, and life insurance problem of a household over a finite time period $[0,T]$, where $T<\infty$ is a constant. Consider a complete probability space $(\Omega, \mathcal{F}, \mathbb{F}, \mathbb{P})$, where $\mathbb{P}$ is a probability measure, and $\mathbb{F}=\{\mathcal{F}_t\}_{t\in [0,T]}$ is a filtration generated by a standard Brownian motion $W(t)$ and a non-homogeneous Markov chain $\eta(t)$. The state space of $\{\eta(t), t\in [0,T]\}$ is given by $\mathbb{S}=\{0,1,2,\cdots,N\}$.

\vspace{0.2cm}

Suppose that the household consists of a breadwinner with income and a dependent without income. $T$ represents the time limit for the breadwinner's maintenance obligation. The non-homogeneous Markov chain $\eta(t)$ is used to describe the breadwinner's health status. When it takes the value 0, it indicates that the breadwinner is healthy. When $\eta(t)$ takes a value in the set $\mathbb{S} \backslash \{0\}$, it represents the breadwinner being in a state of accident, disability, illness, or other health-related state. This can be referred to as a non-healthy state (which encompasses the general health risks including the risk of death, although this study focuses on the impact of life insurance for mitigating the risk of death, and therefore does not include the death state in the health status). The health status $\eta(t)$ is not always $0$, reflecting the risk of the breadwinner experiencing health shocks.
The matrix $Q(t) = (q_{ij}(t)_{\{i,j\in \mathbb{S}\}})$ is the transition intensity matrix, where $q_{ij}(t)$ represents the transition intensity from state $i$ to state $j$ at time $t$.

\subsection{Insurance market}

Assuming that $\tau_z $ is the remaining life of the breadwinner when his/her current state is $z$($ z\in \mathbb{S})$, then $\tau_z$ is a non-negative random variable. Suppose $\tau_z$ has a probability density function $f(t,z)$ and a distribution function $F(t,z) = \mathbb{P}(\tau_z < t) = \int_{0}^{t}f(u,z) {\rm d}u$. Let $\lambda(t,z)$ denote the death force of a breadwinner in state $z$ at time $t$, then
$$\lambda(t,z)=\mathop {\lim }\limits_{\varepsilon \to 0 }  {\frac{\mathbb{P}(t \le \tau_z <t+\varepsilon | \tau_z \ge t)}{\varepsilon}} .$$
Thus, $\bar{F}(t,z)=1-F(t,z)={\rm exp}\{-\int_{0}^{t}\lambda(u,z){\rm d}u\}$, $f(t,z)=\lambda(t,z) {\rm exp}\{-\int_{0}^{t}\lambda(u,z){\rm d}u\}$.

\vspace{0.2cm}

Suppose an insurance company sells an instantaneous (infinitely short period of time) life insurance, which is priced according to the state of the insured, and charges a life insurance rate denoted as $\theta(t,z)$. Taking into account the operating costs of the insurance company, there is usually $\theta(t,z)\ge \lambda(t,z)$. For convenience, if a breadwinner spends $p(t,z)$ at time $t$ on this life insurance, and dies immediately, the compensation received by the household is recorded as $\displaystyle{\frac{p(t,z)}{\lambda(t,z)}}$. That is, the insurance market is frictionless, e.g., Shen and Sherris (2018) , Wang et al. (2021).

\subsection{Consumption habits}

When households make consumption decisions, their current consumption behaviour is influenced by their previous consumption levels. Let $c(t)$ denote the consumption rate at $t$ and $h(t)$ the consumption habits function. According to the model of Detemple and Zapatero (1992), the consumption habits are as follows
\begin{equation}
	h(t)={\rm e}^{-\beta t}h_0+\alpha \int_{0}^{t}{\rm e}^{\beta (s-t)}c(s){\rm d}s,
	\nonumber	
\end{equation}
where $\alpha$, $\beta$ and $h_0$ are constants.
$\alpha$ measures the impact of historical consumption levels on current consumption levels. The larger the $\alpha$, the more important it is for households to maintain their current level of consumption. $\beta$ measures the extent to which households have forgotten about their past consumption. The larger the $\beta$, the less influence the past consumption has on the current consumption levels.

Its differential form is
\begin{equation}
	{\rm d} h(t)=[\alpha c(t)-\beta h(t)]{\rm d} t,	
\end{equation}

\subsection{Financial market}

Assuming that the financial market consists of a risk-free asset and a risky asset. The price of the risk-free asset $S_0(t)$ and the price of the risky asset $S_1(t)$ are given by

\begin{equation}
	{\rm d} S_0(t)=rS_0(t){\rm d} t,
\end{equation}
\begin{equation}
	{\rm d} S_1(t)=S_1(t)[\mu{\rm d}t+\sigma{\rm d}W(t)],	
\end{equation}
where $r$ stands for the interest rate on risk-free assets. $\mu>r$ is the expected return on risky assets. $\sigma$ is the volatility of a risky asset. $W(t)$ is the standard Brownian motion.
\subsection{Wealth process}

The income of a breadwinner depends on his/her health status. Denote the wage rate of a breadwinner in a health state as $y(t,0)$. In the case of unhealthy state, the income is denoted by $y(t,k)=\frac{1}{\xi_k} y(t,0),\ k\in \mathbb{S} \backslash \{0\} $. Where $\xi_k>1$ is a constant related to state and its value reflects the intensity of the health shock. Considering the social welfare and security measures, it is set to $\xi_k<\infty$, i.e., when the breadwinner is in a non-healthy state, the household still has income.

\vspace{0.2cm}

Let $X(t)$ denote the wealth of the household at time $t$. Let $\nu=(\pi(t),c(t),p(t,\eta(t)))$ denote the strategy of the household, where $\pi(t)$ is the amount invested in the risky assets, $c(t)$ is the consumption rate and $p(t,\eta(t))$ is the life insurance premium rate.
Given the strategy $\nu$, the wealth process of the household is as follows
\begin{equation}
		\left\{ {\begin{array}{*{20}{l}}
				{\displaystyle{{\rm d}X(t)=\pi(t) \frac{{\rm d} S_1(t)}{S_1(t)}+[X(t)-\pi(t)] \frac{{\rm d} S_0(t)}{S_0(t)}+\mathbb{I}_{\{t<\tau_{\eta(t)}\}}[y(t,\eta(t))-p(t,\eta(t))]{\rm d}t-c(t){\rm d} t,}}\\
				{X(0) = x_0,\eta(0)=\eta_0.}
		\end{array}} \right.
	\end{equation}	

Thus,
\begin{equation}
		\left\{ {\begin{array}{*{20}{l}}
				{\displaystyle{{\rm d}X(t)=\{rX(t)+\pi(t)(\mu-r)+\mathbb{I}_{\{t<\tau_{\eta(t)}\} }[y(t,\eta(t))-p(t,\eta(t))]-c(t)\}{\rm d} t+\sigma \pi(t){\rm d}W(t),}}\\
				{X(0) = x_0,\eta(0)=\eta_0.}
		\end{array}} \right.
	\end{equation}

\begin{Definition}(Admissible Strategy).

For any $t\in [0,T]$, a strategy $\nu=(\pi(s),c(s),p(s,\eta(s)))_{s\in[t,T]}$ is admissible if

(i) $\nu$ is $\mathcal{F}_t$ - adaptive;

(ii) For any $s \in [t,T], c(t)\ge 0$ and $\displaystyle{E\left[\int_{t}^{T}[\pi(s)^2+c(s)^2+p(s,\eta(s))^2]{\rm d}s\right]}<+\infty;$

(iii) $(X^{\nu},\nu)$ is the unique solution of the equation $(2.5)$.

The admissible set is expressed as $\Pi$.

\end{Definition}

\vspace{0.2cm}

The objective function is
\begin{equation}
		J(t,x,h,i;\nu(\cdot))=E_{t,x,h,i}\left[\int_{t}^{T}U(s,c(s),h(s),\eta(s)){\rm d}s+\Psi(T,X(T),\eta(T))\right],
	\end{equation}
where $U(t,c,h,i)$ denotes the consumption utility of a household with a habit level of $h$ in state $i$ at time $t$, $\Psi(T,X(T),\eta(T))$ denotes the utility function of terminal wealth.

\vspace{0.2cm}

Referring to Tao et al. (2023), the value function can be defined as
\begin{equation}
		\left\{ {\begin{array}{*{20}{l}}
				{\displaystyle{V(t,x,h,i)=\mathop{{\rm sup}}\limits_{{\nu}\in \Pi}J(t,x,h,i;\nu(\cdot)),}}\\
				{V(T,x,h,i)=\Psi(T,x,i), \quad ((t,x,h,i)\in [0,T]\times \mathbb{R} \times \mathbb{R} \times \mathbb{S}).}
		\end{array}} \right.
	\end{equation}

\vspace{0.2cm}

The optimal strategy is $\nu^{*}(t)=(\pi^{*}(t),c^{*}(t),p^{*}(t,i)).$

\section{Determination of optimal solutions}

\subsection{Optimization problem after the death of the breadwinner}

If the breadwinner dies before $T$, there is no change of state afterwards.
When $t\in [\tau_i ,T]$, dependent use wealth for investment and consumption.
The strategy is $\nu_d(t)=(\pi_d(t),c_d(t))$, where $\pi_d(t)$ is the amount invested in risky assets and $c_d(t)$ is consumption rate. The corresponding set of admissible is denoted as $\Pi_d$.
In this stage, the objective function is
\begin{equation}
	J_d(t,x_d,h_d;\nu_d(\cdot))=E_{t,x_d,h_d}\left[\int_{t}^{T}U_d(s,c_d(s),h_d(s)){\rm d}s+\Psi_d(T,X_d(T))\right],
\end{equation}
where, $X_d(t)$ is the household wealth process under the strategy $\nu_b$.
\begin{equation}
\displaystyle{{\rm d}X_d(t)=[rX_d(t)+\pi_d(t)(\mu-r)-c_d(t)]{\rm d}t+\sigma\pi_d(t){\rm d}W(t).}
\end{equation}
$h_d(t)$ denotes the consumption habits levels, which satisfying
\begin{equation}
	\displaystyle{{\rm d}h_d(t)=[\alpha c_d(t)-\beta h_d(t)]{\rm d}t.}
\end{equation}

\vspace{0.2cm}

The value function corresponding to $(3.1)$ is well-defined by
\begin{equation}
    \left\{ {\begin{array}{*{20}{l}}
	{\displaystyle{V_d(t,x_d,h_d)=\mathop{{\rm sup}}\limits_{{\nu_d}\in \Pi_d}J_d(t,x_d,h_d;\nu_d(\cdot)),}}\\
    {\displaystyle{V_d(T,x_d,h_d)=\Psi(T,x_d).}}
    \end{array}} \right.
\end{equation}

\vspace{0.2cm}

Suppose the utility functions as
$$U_d(t,c_d,h_d)=k_d {\rm e}^{-\rho t}\frac{(c_d-h_d)^{1-\gamma}}{1-\gamma},$$
$$\Psi_d(t,x_d)=\omega_d {\rm e}^{-\rho t}\frac{x_d^{1-\gamma}}{1-\gamma}.$$
$\rho$ is the discount rate, $\gamma$ is the risk preference parameter, $k_d$, $\omega_d$ is the weight coefficient.
Note that consumption habits can be regarded as the lowest level of consumption and only consumption beyond the basic needs of life can make people satisfied.
Therefore, instead of considering the utility of all consumption, we consider the utility of consumption above the consumption habit part.
This means that consumption to remain above the habit level.

\vspace{0.2cm}

Using the dynamic programming approach, the corresponding HJB equation of optimization problem $(3.4)$ is given by

\begin{equation}
	\mathop{{\rm sup}}\limits_{{\nu}_d\in\Pi_d}\left\{U_d(t,c_d,h_d)+V_{d,t}+[rx_d+\pi_d(\mu-r)-c_d]V_{d,x}+(\alpha c_d-\beta h_d)V_{d,h}+\frac{1}{2}\sigma^2\pi^2_dV_{d,xx}\right\}=0,
\end{equation}
where $V_{d, t}, V_{d, x}$ and $V_{d, h}$ denote the first-order partial derivatives of $V_d$ with respect to $t, x$, $h$, $V_{d, xx}$ represent the second-order partial derivatives of $V_d$ with respect to $x$.

\vspace{0.3cm}

\begin{Theorem}(Verification Theorem).
Let $V_d(t, x_d, h_d)$ be a solution of the HJB
equation. Then, the inequality $$V_d(t, x_d, h_d) \ge J_d(t, x_d, h_d;\nu_d(\cdot))$$
holds for every $\nu_d(\cdot) \in \Pi_d $ and $(t, x_d, h_d) \in [0, T ) \times \mathbb{R} \times \mathbb{R}$. Furthermore,
an admissible pair $(X_d^*(t), h^*_d(t), \pi_d^*(t), c_d^*(t))$ is optimal if and only if the equality

\begin{equation}
		\begin{aligned}
&U_d(t,c_d^*(t),h^*_d(t))+V_{d,t}+[rX_d^*(t)+\pi_d^*(t)(\mu-r)-c_d^*(t)]V_{d,x}\\
&+(\alpha c_d^*(t)
-\beta h^*_d(t))V_{d,h}+\frac{1}{2}\sigma^2(\pi^*_d(t))^2 V_{d,xx}=0,
    \nonumber
		\end{aligned}
\end{equation}
holds for $a.e. t \in [s, T ] $ and $\mathbb{P} - a.s.$
\end{Theorem}

\begin{Theorem} The value function of the optimal control problem $(3.4)$ is
\begin{equation}
	\displaystyle{V_d(t,x_d,h_d)=\frac{1}{1-\gamma}[g(t)]^{\gamma}[x_d-h_dB(t)]^{1-\gamma}.}
\end{equation}

The corresponding optimal strategy is
\begin{equation}
	\left\{ {\begin{array}{*{20}{l}}
		{\pi_d^*(t)=\displaystyle{\frac{\mu - r}{\sigma^2 \gamma}[x_d-h_d B(t)]},}\\
		{c_d^*(t)=\displaystyle{h_d+\frac{x_d-h_d B(t)}{g(t)}[1+\alpha B(t)]^{-\frac{1}{\gamma}}(k_d {\rm e}^{-\rho t})^{\frac{1}{\gamma}},}}
	\end{array}} \right.
\end{equation}
where
\begin{equation}
			\begin{aligned}
				B(t) &= \frac{1}{r+\beta-\alpha}[1-{\rm e}^{-(r+\beta-\alpha)(T-t)}],\\	
				g(t) &= \int_{t}^{T}{\rm e}^{N(s-t)}(k_d {\rm e}^{-\rho s})^{\frac{1}{\gamma}}[1+\alpha B(s)]^{1-\frac{1}{\gamma}} {\rm d}s+{\rm e}^{N(T-t)}(\omega_d {\rm e}^{-\rho T})^{\frac{1}{\gamma}},\\
				N &= \frac{1-\gamma}{\gamma}[r+\frac{(\mu -r)^2}{2\sigma^2 \gamma}].
			\end{aligned}	
		\end{equation}
\end{Theorem}

\noindent
See Appendix for proof.

\subsection{Optimization problem when the breadwinner is alive}

When $t\in [0,\tau_i \land T]$, the household needs to make the optimal investment, consumption and life insurance decisions.
The strategy is $\nu_a(t)=(\pi_a(t),c_a(t),p_a(t,i))$, where $\pi_a(t)$ is the amount invested in risky assets, $c_a(t)$ represents the consumption rate of the household, $p_a(t,i)$ represents the life insurance premiums rate, and the corresponding set of permissible strategies is denoted as $\Pi_a$.
The objective function $(2.6)$ of the household can be rewritten as
\begin{equation}
		\begin{aligned}	
			&J(t,x_a,h_a,i;\nu_a(\cdot)) \\
			=&E_{t,x_a,h_a,i}\Big[\int_{t}^{\tau_i \land T}U_a(s,c_a(s),h_a(s),\eta(s)){\rm d}s+\mathbb{I}_{\{\tau_i > T \}}\Psi_a(T,X_a(T),\eta(T))\\
			&+\mathbb{I}_{\{\tau_i < T \}}[\int_{\tau_i}^{T}U_d(s,c_d(s),h_d(s)){\rm d}s+\Psi_d(T,X_d(T))]\Big],
		\end{aligned}	
	\end{equation}	
where $X_a(t)$ is wealth process under the strategy $\nu_a$.
\begin{equation}
		\displaystyle{{\rm d}X_a(t)=[rX_a(t)+\pi_a(t)(\mu-r)+y(t,i)-p_a(t,i)-c_a(t)]{\rm d}t+\sigma\pi_a(t){\rm d}W(t).}
	\end{equation}

\noindent
$h_a(t)$ represents the level of consumption habits of the household,
\begin{equation}
	\displaystyle{{\rm d}h_a(t)=[\alpha c_a(t)-\beta h_a(t)]{\rm d}t.}
\end{equation}

\vspace{0.2cm}

According to Lemma 3.3.1 of Ye (2006), the following Lemma is obtained.

\begin{Lemma}
The objective function $(3.9)$ can be expressed as
\begin{equation}
			\begin{aligned}	
				&J(t,x_a,h_a,i;\nu_a(\cdot))\\
				=&\frac{1}{\bar{F}(t,i)}E_{t,x_a,h_a,i}\Big\{\int_{t}^{T}\left[\bar{F}(s,\eta(s))U_a(s,c_a(s),h_a(s),\eta(s))
+f(s,\eta(s))V_d(s,x_a(s) \right. \\
&\left.+\frac{p_a(s,\eta(s))}{\lambda(s,\eta(s))},h_d(s))\right]{\rm d}s +\bar{F}(T,\eta(T))\Psi_a(T,X_a(T),\eta(T))\Big\}.
			\end{aligned}	
		\end{equation}
\end{Lemma}

\noindent
See Appendix for proof.

\vspace{0.2cm}

Suppose
\begin{equation}
		\tilde{J}(t,x_a,h_a,i; \nu_a(\cdot))=\bar{F}(t,i)J(t,x_a,h_a,i;\nu_a(\cdot)).
\end{equation}
Then, the corresponding optimal control problem is given
\begin{equation}
\left\{ {\begin{array}{*{20}{l}}		
		{\displaystyle{\tilde{V}(t,x_a,h_a,i)=\mathop{{\rm sup}}\limits_{{\nu_a}\in \Pi_a}\tilde{J}(t,x_a,h_a,i;\nu_a(\cdot)),}}\\
    {\displaystyle{\tilde{V}(T,x_a,h_a,i)=\bar{F}(T,i)\Psi_a(T,x_a,i).}}
\end{array}} \right.
\end{equation}

\vspace{0.2cm}

Consider the utility function as
$$U_a(t,c_a,h_a,i)=k_{a,i} {\rm e}^{-\rho t}\frac{(c_a-h_a)^{1-\gamma}}{1-\gamma},$$
$$\Psi_a(t,x_a,i)=\omega_{a,i} {\rm e}^{-\rho t}\frac{x_a^{1-\gamma}}{1-\gamma}.$$
$\rho$ is the discount rate, $\gamma$ is the risk preference parameter, and the effect of state $i$ is reflected by $k_{a,i}, \omega_{a,i}$.

\vspace{0.2cm}

Referring to theorem 2.4 of Tao et al. (2023), the HJB equation is given.
\begin{equation}
		\tilde{V}_{t}+\mathop{{\rm sup}}\limits_{{\nu}_a\in\Pi_a} \mathcal{H}(t, x_a, h_a, i;\nu_a(\cdot))=0.
	\end{equation}

\begin{equation}
		\begin{aligned}	
			\mathcal{H}(t, x_a, h_a, i;\nu_a(\cdot))=&\sum\limits_{j\in \mathbb{S};j\ne i} {q_{ij}(t)[\tilde{V}(t,x_a,h_a,j)-\tilde{V}(t,x_a,h_a,i)]}+[rx_a+\pi_a(\mu-r)+y(t,i) \\
			&-p_a(t,i)-c_a]\tilde{V}_{x} +(\alpha c_a-\beta h_a)\tilde{V}_{h}+\bar{F}(t,i)U_a(t,c_a,h_a,i) \\
			&+\frac{1}{2}\sigma^2\pi^2_a\tilde{V}_{xx}+f(t,i)V_d(t,x_a(t)+\frac{p_a(t,i)}{\lambda(t,i)},h_d)	
		\end{aligned}
	\end{equation}
where $\tilde{V}_{t},\tilde{V}_{x}$ and $\tilde{V}_{h}$ denote the first-order partial derivatives of $\tilde{V}$ with respect to $t, x$, $h$, $\tilde{V}_{xx}$ represent the second-order partial derivative of $\tilde{V}$ with respect to $x$.

\vspace{0.2cm}

\begin{Theorem}(Verification Theorem).
Let $\tilde{V}(t, x_a, h_a, i)$ be a solution of the HJB
equation. Then, the inequality $$\tilde{V}(t, x_a, h_a, i) \ge \tilde{J}(t, x_a, h_a, i;\nu_a(\cdot))$$
holds for every $\nu_a(\cdot) \in \Pi_a $ and $(t, x_a, h_a, i) \in [0, T ) \times \mathbb{R} \times \mathbb{R} \times \mathbb{S} $. Furthermore,
an admissible pair $(X_a^*(t), h^*_a(t), \nu_a^*(t))$ is optimal if and only if the equality
$$\tilde{V}_{t}+ \mathcal{H}(t, X^*_a(t), h^*_a(t),\eta(t);\nu_a^*(t))=0, $$
holds for $a.e. t \in [s, T ] $ and $\mathbb{P} - a.s.$
\end{Theorem}

\begin{Theorem} The value function of the optimal control problem $(3.14)$ is
\begin{equation}
		\displaystyle{\tilde{V}(t,x_a,h_a,i)=\frac{1}{1-\gamma}[G_i(t)]^{\gamma}[x_a+M(t)+h_a A(t)]^{1-\gamma}.}
	\end{equation}

The corresponding optimal strategy is $$\nu_a^*(t)=(\pi_a^*(t),c_a^*(t),p_a^*(t,i))
    =(\pi_a^*(t,X_a^*(t),\eta(t)),c_a^*(t,X_a^*(t),\eta(t)),p_a^*(t,X_a^*(t),\eta(t))).$$
\begin{equation}
		\left\{ {\begin{array}{*{20}{l}}
				{\pi_a^*(t,x_a,i)=\displaystyle{\frac{\mu - r}{\sigma^2 \gamma}[x_a+M(t)+h_a A(t)]},}\\
				{c_a^*(t,x_a,i)=\displaystyle{h_a+\frac{x_a+M(t)+h_a A(t)}{G_i(t)}[1-\alpha A(t)]^{-\frac{1}{\gamma}}[k_{a,i} \bar{F}(t,i) {\rm e}^{-\rho t}]^{\frac{1}{\gamma}},}}\\
				{p_a^*(t,x_a,i)=\displaystyle{\lambda(t,i)\Big[h_dB(t)-x_a+\frac{x_a+M(t)+h_a A(t)}{G_i(t)}[\lambda(t,i)]^{-\frac{1}{\gamma}}[f(t,i)]^{\frac{1}{\gamma}}g(t)\Big]},}
		\end{array}} \right.
	\end{equation}
where
\begin{equation}
    \begin{aligned}
	A(t) &= \displaystyle{-\int_{t}^{T}{\rm e}^{-\int_{t}^{s}[r+\lambda(u,\eta(u))+\beta-\alpha]{\rm d}u}{\rm d}s,} \\
	M(t) &= \displaystyle{\int_{t}^{T}{\rm e}^{-\int_{t}^{s}[r+\lambda(u,\eta(u))]{\rm d}u}[y(s,\eta(s)) -\lambda(s,\eta(s))h_dB(s)]{\rm d}s,}
    \end{aligned}	
\end{equation}
$G_i(t)$ satisfies a system of differential equations
\begin{equation}
	\left\{ {\begin{array}{*{20}{l}}				{G^{'}_i(t)+\frac{1}{\gamma}\Big[(1-\gamma)[r+\lambda(t,i)+\frac{(\mu-r)^2}{2\sigma^2\gamma}]+\sum\limits_{j\in \mathbb{S};j\ne i} {q_{ij}(t)[(\frac{G_j(t)}{G_i(t)})^{\gamma}-1]} \Big]G_i(t)} \\
	{\quad +[k_{a,i}\bar{F}(t,i){\rm e}^{-\rho t}]^{\frac{1}{\gamma}}[1-\alpha A(t)]^{1-\frac{1}{\gamma}}+[f(t,i)]^{\frac{1}{\gamma}}[\lambda(t,i)]^{1-\frac{1}{\gamma}}g(t)=0,}\\
	{G_i(T)=[\omega_{a,i}\bar{F}(T,i){\rm e}^{-\rho T}]^{\frac{1}{\gamma}}.}
   \end{array}} \right.
\end{equation}
\end{Theorem}

\noindent
See Appendix for proof.

\vspace{0.2cm}

According to equation $(2.7)$, $(3.13)$ and $(3.14)$, $\tilde{V}(t,x_a,h_a,i)=\bar{F}(t,i)V(t,x_a,h_a,i)$, then the optimal value function of the optimal control problem $(2.7)$ is
\begin{equation}
\displaystyle{V(t,x_a,h_a,i)=\frac{1}{\bar{F}(t,i)}\tilde{V}(t,x_a,h_a,i)=\frac{1}{(1-\gamma)\bar{F}(t,i)}
    [G_i(t)]^{\gamma}[x_a+M(t)+h_a A(t)]^{1-\gamma}.}
\end{equation}
and the optimal strategy is $\nu_a^*(t)=(\pi_a^*(t,X_a^*(t),\eta(t)),c_a^*(t,X_a^*(t),\eta(t)),p_a^*(t,X_a^*(t),\eta(t))).$

\section{Calibration}

The model established in this paper is applicable to many scenarios. For convenience, we set $\mathbb{S}=\{0,1\}$.
When $q_{10}(t) = 0$ is constant, state 1 is the absorption state, which can be used to represent the permanent disability model.
When the unhealthy state 1 is injury, it can represent the accidental injury model.
Taking state 1 as disease, the influence of disease on optimal household investment, consumption and life insurance strategies can be explored.
In addition, it can also be used to study household asset allocation under the impact of unemployment risk.
Here, the non-healthy state 1 is taken as critical illness, and the permanent health shock is considered by referring to the Markov model of Hambel(2020).
Failure of the breadwinner to recover, as shown in Figure 4.1.
Health shock can be interpreted as a transition from a healthy to a critical illness of the breadwinner.

\begin{figure}[htp!]
	\centering
	\includegraphics[width=.4\textwidth]{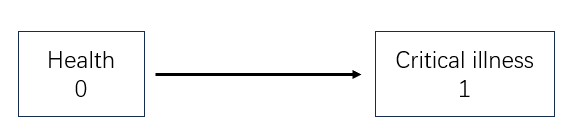}
	\caption{Critical illness model.}
\end{figure}

Assuming that the mortality rate in healthy breadwinner follows the Gompertz mortality model,
\begin{equation}
	\lambda(t,0)=\displaystyle{\frac{1}{n}{\rm exp}\{\frac{m+t-l}{n}\}. }	
\end{equation}
We calibrated the parameters based on data from China Life Insurance Mortality Table (2010-2013).
The table of non-elderly care business aged 20-60 years old is selected, and the weighted average of the data of the same age and different genders is used as the real data.
Gender is not differentiated in the numerical simulation, and Figure 4.2 compares simulated mortality with empirical data.
According to the fitting curve, we set the parameters $m = 20, n =12.14982, l = 92.29736$.

\begin{figure}[htbp]
	\centering
	\begin{minipage}[c]{0.48\textwidth}
		\centering
		\includegraphics[width=\textwidth]{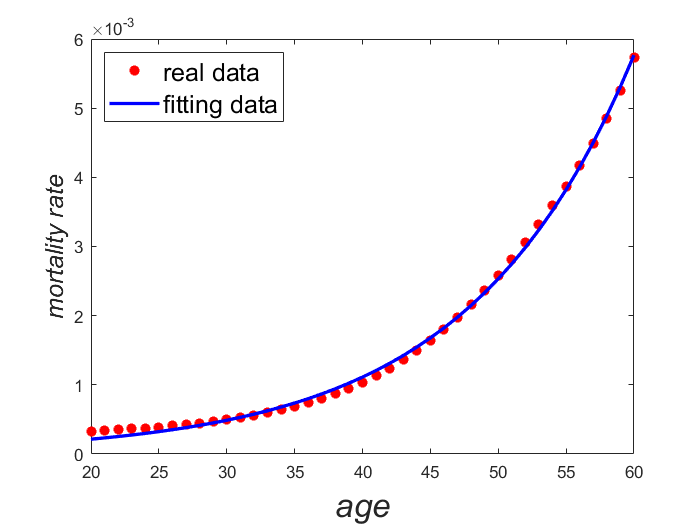}
		\caption{Mortality model calibration.}
	\end{minipage}
	\begin{minipage}[c]{0.48\textwidth}
		\centering
		\includegraphics[width=\textwidth]{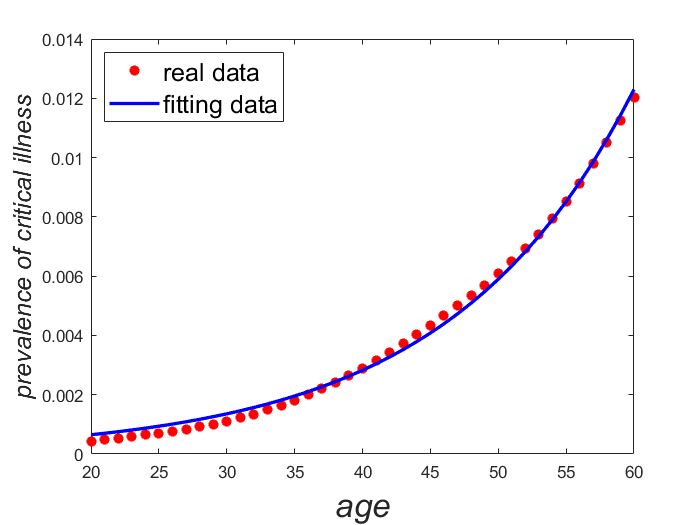}
		\caption{Critical illness prevalence calibration.}
	\end{minipage}
\end{figure}

In addition, according to the China Life Insurance Critical Illness Morbidity Table (2020), the proportion of serious diseases deaths continues to rise throughout the middle age, especially between the ages of 40 and 75, the proportion of serious diseases deaths exceeds 50$\%$.
Therefore, we have reason to believe that the occurrence of critical illness increases the risk of death, and the increase in mortality is also related to age.
Suppose that the mortality rate of the breadwinner with a critical illness follows the pattern
\begin{equation}
	\lambda(t,1)=\displaystyle{\frac{1}{n}{\rm exp}\{\frac{m+t-l}{n}\}+k_1+k_2(m+t). }	
\end{equation}
Combined with mortality data and critical illness data, the calibration results are $k_1=0.032, k_2=0.0043$.

Suppose that the probability of a healthy breadwinner suffering a critical illness follows the exponential form, i.e. $q_{01}(t)=m_1{\rm e}^{n_1t}$.
Similarly, the parameters were calibrated using the China Life Insurance Critical Illness Morbidity Table (2020).
The fitting results are shown in Figure 4.3. Parameters $m_1 =0.0001492$ and $n_1 = 0.07353$ are set.

\section{Sensitivity analysis}

This section takes the critical illness model as an example to discuss the influence of important parameters on optimal household investment, consumption and life insurance decisions based on the current physical state of the breadwinner.

The following assumptions are made about parameter values in this paper.

$ \bullet $ The age from 20 to 60 is the period when the breadwinners take on household responsibilities, so we take $T=40$.

$ \bullet $ According to Boyle et al. (2022), the initial wealth value of a household consisting of one breadwinner and one dependent is $x_0 = 35,000$, the initial consumption habits level is $h_0=6$, the risk aversion parameter is $\gamma=6$, and the financial market parameters are $r=0.02, \mu=0.07, \sigma=0.2$.

$ \bullet $ Based on Kraft et al. (2017), the consumption habits parameters were set as $\alpha=0.1$ and $\beta =0.174$.

$ \bullet $ We set the discount rate $\rho=0.1$, which is consistent with the empirical studies of Andersen et al. (2008) and Love (2009).

$ \bullet $ According to Tao et al. (2023), the weight coefficient of household cumulative consumption utility is $ k_{a,0} =1, k_{a,1} =0.5, k_d =0.5 $, and the weight coefficient of household terminal wealth utility is set as $ \omega_{a,0} = 2.5, \omega_{a,1} = 3, \omega_d = 3 $.

$ \bullet $ Suppose that the labor income of the breadwinner grows exponentially, that is, $y(t,i) = y(0,i){\rm e}^{\delta t}$.
Referring to the empirical data of Liu et al. (2020), $\delta =0.075$.
In addition, according to Hambel (2020), the initial income of a healthy breadwinner is $y(0,0) = 25,000$, and the income decreases by 20$\%$ after suffering a health shock, so $\xi_1 =1.25$.

\subsection{The impact of health shock on households optimal investment, consumption, and life insurance strategies}

\begin{figure}
	\centering
	\subfigure[]{
		\begin{minipage}[t]{0.32\textwidth}
			\centering
			\includegraphics[width=\textwidth]{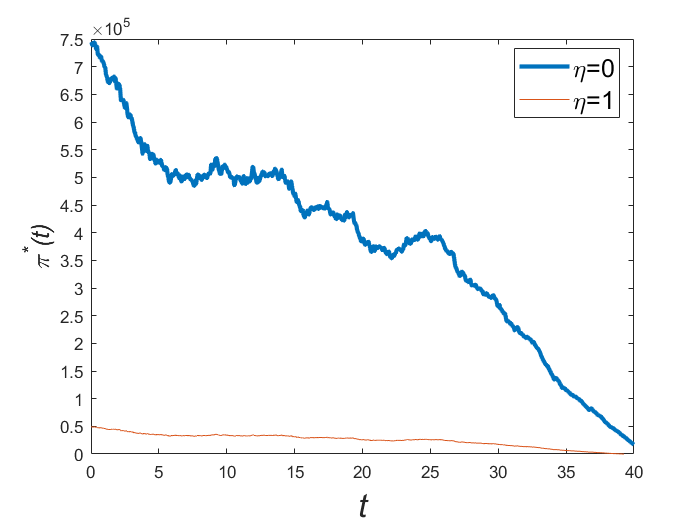}\\
			\vspace{0.02cm}
			
		\end{minipage}%
	}%
	\subfigure[]{
		\begin{minipage}[t]{0.32\textwidth}
			\centering
			\includegraphics[width=\textwidth]{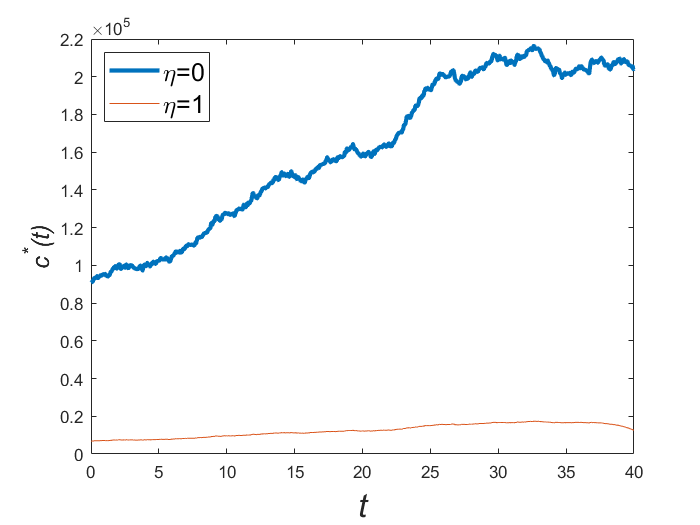}\\
			\vspace{0.02cm}
			
		\end{minipage}%
	}%
	\subfigure[]{
		\begin{minipage}[t]{0.32\textwidth}
			\centering
			\includegraphics[width=\textwidth]{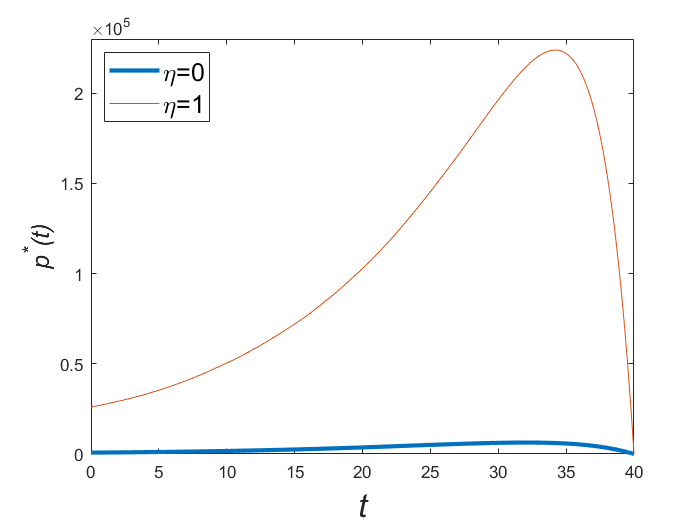}\\
			\vspace{0.02cm}
			
		\end{minipage}%
	}%
	\centering
	\caption{The influence of the physical state of the breadwinner $\eta$ on the household investment strategy $\pi^*$, consumption strategy $c^*$, and life insurance strategy $p^*$. }
	\vspace{-0.2cm}
\end{figure}
	
Figure 5.1 shows the influence of the physical state of the breadwinner on the optimal investment, consumption and life insurance strategies of the household.
According to Figure 5.1 (a), households facing critical illnesses tend to hold lower-risk assets. This is due to the increased healthcare expenses and decreased household wealth caused by the breadwinner's illness, leaving households fewer able to cope with financial risks. Consequently, these households prefer to invest their funds in risk-free assets.
Figure 5.1 (b) depicts the impact of the physical state of the breadwinner on household consumption.
The graph reveals that households with a healthy breadwinner have higher consumption expenditures compared to those with an ill breadwinner.
Due to the increase in medical expenditure, the consumption expenditure of households with ill breadwinner will decrease correspondingly when the total amount of resources is given.
Figure 5.1 (c) shows the impact of the physical state on the life insurance strategy. The figure shows that the effect of the physical state of the breadwinner on the life insurance strategy is significant. Among households with healthy breadwinners, the life insurance expenditure curve is smoother and does not fluctuate substantially. In general, breadwinners in good health have a longer life expectancy and a relatively stable household financial situation. Therefore, they can choose a relatively stable life insurance strategy to ensure the economic security of the households after their death. There is a large fluctuation in the life insurance strategy among households whose breadwinners have a medical condition.
	
\begin{figure}
	\centering
	\subfigure[]{
		\begin{minipage}[t]{0.32\textwidth}
			\centering
			\includegraphics[width=\textwidth]{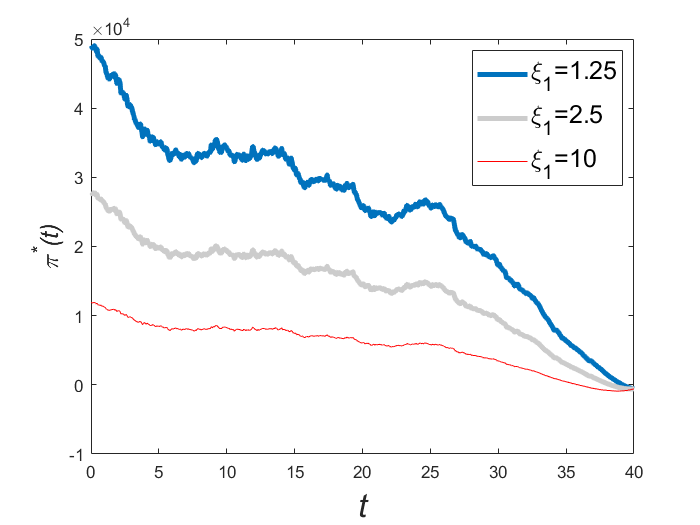}\\
			\vspace{0.02cm}
			
		\end{minipage}%
	}%
	\subfigure[]{
		\begin{minipage}[t]{0.32\textwidth}
			\centering
			\includegraphics[width=\textwidth]{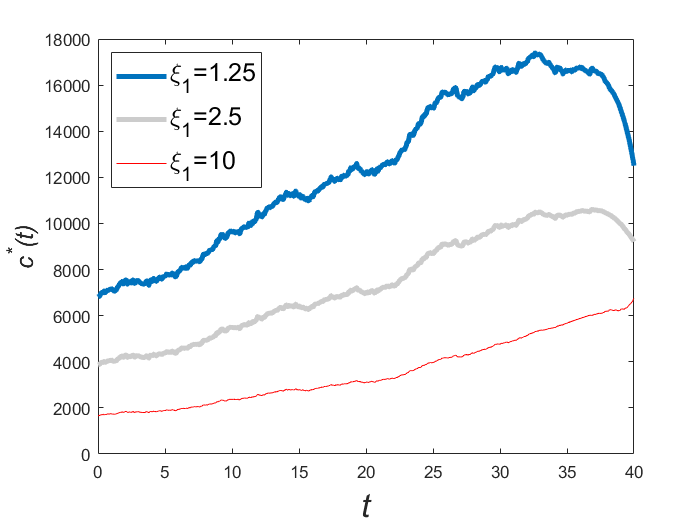}\\
			\vspace{0.02cm}
			
		\end{minipage}%
	}%
	\subfigure[]{
		\begin{minipage}[t]{0.32\textwidth}
			\centering
			\includegraphics[width=\textwidth]{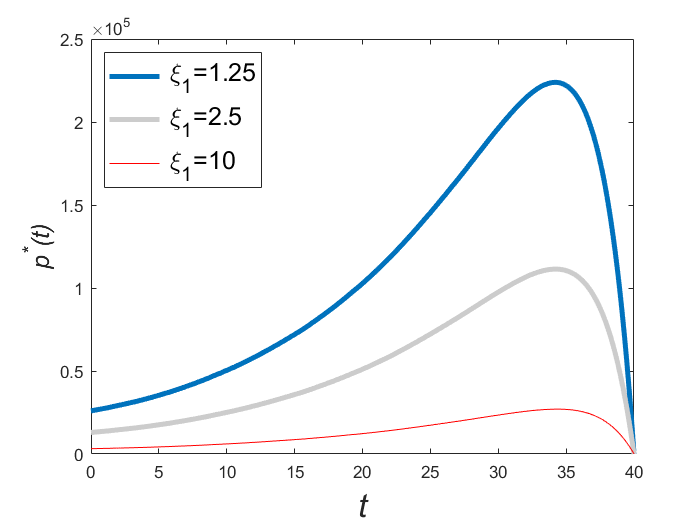}\\
			\vspace{0.02cm}
			
		\end{minipage}%
	}%
	\centering
	\caption{ The influence of breadwinner income parameter $\xi_1$ on household investment strategy $\pi^*$, consumption strategy $c^*$, and life insurance strategy $p^*$. }
	\vspace{-0.2cm}
\end{figure}	

Figure 5.2 describes the influence of parameter $\xi_1$ on the optimal strategy for households.
It can be seen from Figure 5.2 (a) that risky asset investment decreases with the increase of parameter $\xi_1$.
The parameter $\xi_1$ reflects the magnitude of health shock.
A higher value of $\xi_1$ indicates poor health of the breadwinner, and the corresponding household income may be lower. In response to this uncertainty, households tend to reduce their investments in risky assets in order to protect household wealth from further losses.
As a result, when households face health shocks, they should consider adjusting their investment strategies and adopt a relatively conservative investment approach to reduce risk exposure and protect the financial stability.
Likewise, Figure 5.2 (b) reveals a decrease in household consumption expenditure as parameter $\xi_1$ increases.
Households need to adjust their consumption levels to new economic conditions under health shocks.
By reducing consumer spending, households can better manage limited resources.
Additionally, Figure 5.2 (c) indicates a decrease in the amount of life insurance as parameter $\xi_1$ increases.
When the health of the breadwinner is poor, the income is lower, and the sudden death of the breadwinner may have less financial impact on the household, so the corresponding demand for life insurance is reduced.
Households prioritize their current economic woes and use their limited funds for more pressing needs.
As a result, households that are less hit by health shocks may appropriately increase investment, consumption and life insurance spending to make better use of their available resources and boost the economic situation.

	\begin{figure}
	\centering
	\subfigure[]{
		\begin{minipage}[t]{0.32\textwidth}
			\centering
			\includegraphics[width=\textwidth]{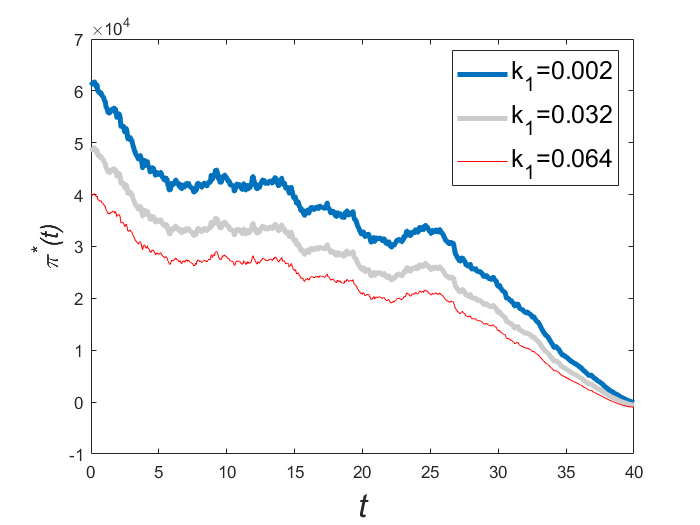}\\
			\vspace{0.02cm}
			
		\end{minipage}%
	}%
	\subfigure[]{
		\begin{minipage}[t]{0.32\textwidth}
			\centering
			\includegraphics[width=\textwidth]{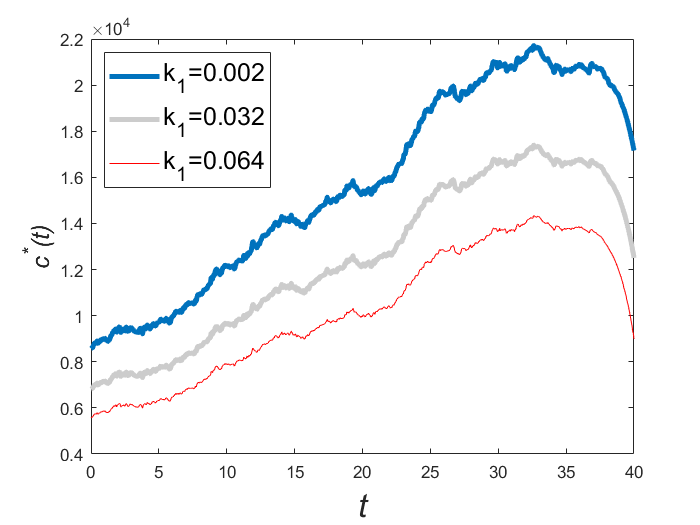}\\
			\vspace{0.02cm}
			
		\end{minipage}%
	}%
	\subfigure[]{
		\begin{minipage}[t]{0.32\textwidth}
			\centering
			\includegraphics[width=\textwidth]{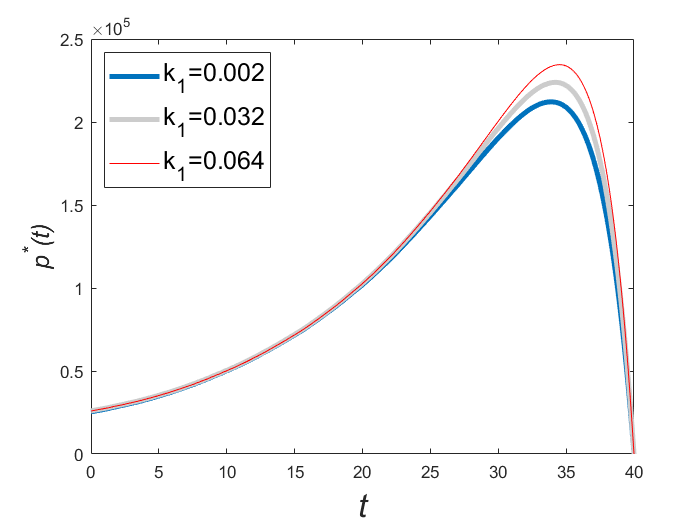}\\
			\vspace{0.02cm}
			
		\end{minipage}%
	}%
	\centering
	\caption{ Influence of breadwinner mortality parameter $k_1$ on household investment strategy $\pi^*$, consumption strategy $c^*$, and life insurance strategy $p^*$.}
	\vspace{-0.2cm}
\end{figure}

Figure 5.3 depicts the influence of parameter $k_1$ on the optimal strategies.
According to the mortality model (4.2) given earlier, $k_1$ measures the fixed increase in mortality from a healthy state to a disease state.
The larger the $k_1$, the more significant the increase in mortality due to health shocks.
As can be seen from Figure 5.3 (a), as the value of $k_1$ increases, households tend to reduce their investment in risky assets.
This is due to the increase in mortality in disease states caused by health shocks, and in order to avoid potential financial losses, households reduce their investments in risky assets.
Similarly, from Figure 5.3 (b), it can be seen that as $k_1$ value increases, household consumption expenditure decreases.
Increased mortality of breadwinners in disease state, exposing households to higher health risks and economic uncertainty.
In order to adapt to this new economic situation, households will adjust consumption levels and reduce consumer spending.
According to Figure 5.3 (c), as mortality increases, so does household expenditure on life insurance.
This is because life insurance provides financial security in the event of the death of a family member and is used for income replacement.
Therefore, in the face of high mortality, households will increase the demand for life insurance and correspondingly increase the life insurance expenditure.

\begin{figure}
	\centering
	\subfigure[]{
		\begin{minipage}[t]{0.32\textwidth}
			\centering
			\includegraphics[width=\textwidth]{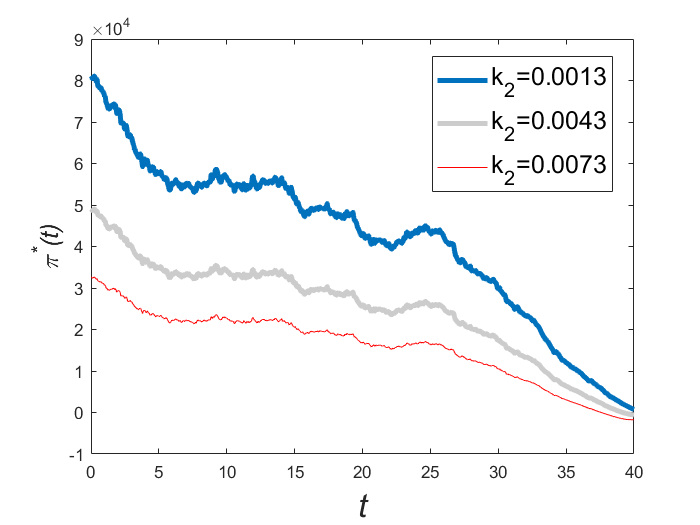}\\
			\vspace{0.02cm}
			
		\end{minipage}%
	}%
	\subfigure[]{
		\begin{minipage}[t]{0.32\textwidth}
			\centering
			\includegraphics[width=\textwidth]{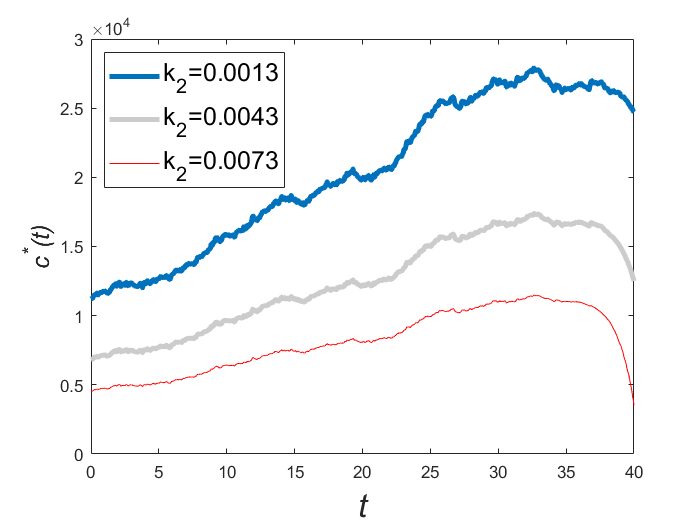}\\
			\vspace{0.02cm}
			
		\end{minipage}%
	}%
	\subfigure[]{
		\begin{minipage}[t]{0.32\textwidth}
			\centering
			\includegraphics[width=\textwidth]{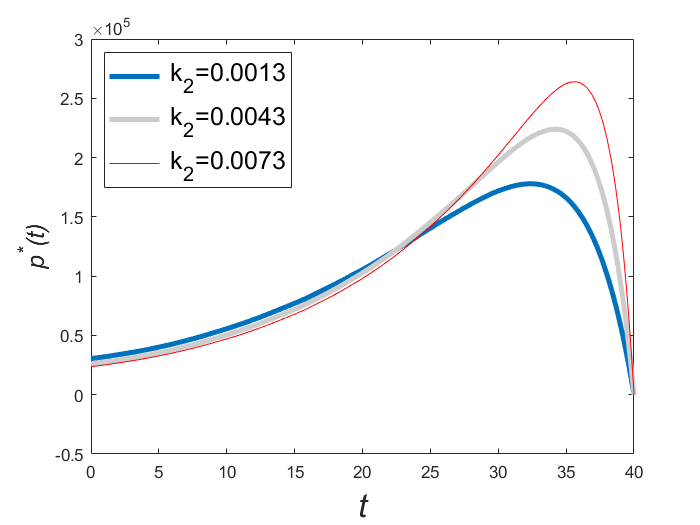}\\
			\vspace{0.02cm}
			
		\end{minipage}%
	}%
	\centering
	\caption{ Influence of breadwinner mortality parameter $k_2$ on household investment strategy $\pi^*$, consumption strategy $c^*$, and life insurance strategy $p^*$. }
	\vspace{-0.2cm}
\end{figure}

Figure 5.4 depicts the influence of parameter $k_2$ on the optimal strategies.
According to the mortality model in Section 4 , $k_2$ measures the rate at which mortality increases with age in disease state.
The higher the value of $k_2$, the faster the mortality rate increases with age, thus the risk investment and consumption expenditure of the household will decrease, and the demand of life insurance will increase.
These strategic adjustments are designed to address the risks and uncertainties associated with increasing mortality rates with age in disease state.
Conversely, households that are less affected by health shocks may appropriately increase investment and consumption to make better use of their available resources and improve the households' economic situation.

\subsection{The impact of consumption habits on optimal household investment, consumption and life insurance strategies}

\begin{figure}
	\centering
	\subfigure[]{
		\begin{minipage}[t]{0.32\textwidth}
			\centering
			\includegraphics[width=\textwidth]{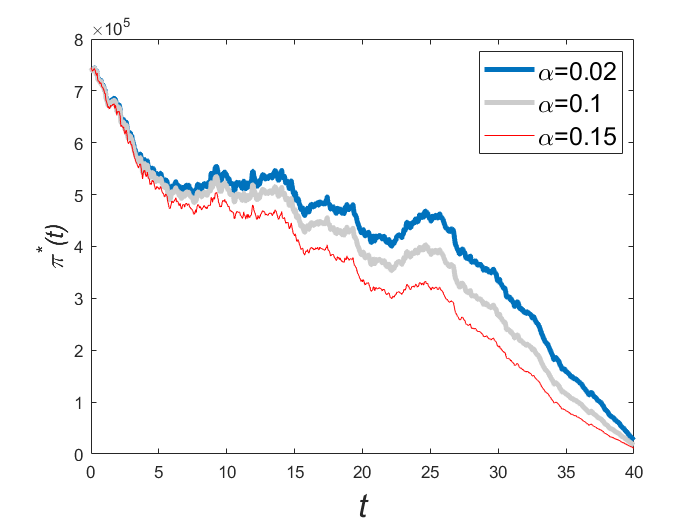}\\
			\vspace{0.02cm}
			
		\end{minipage}%
	}%
	\subfigure[]{
		\begin{minipage}[t]{0.32\textwidth}
			\centering
			\includegraphics[width=\textwidth]{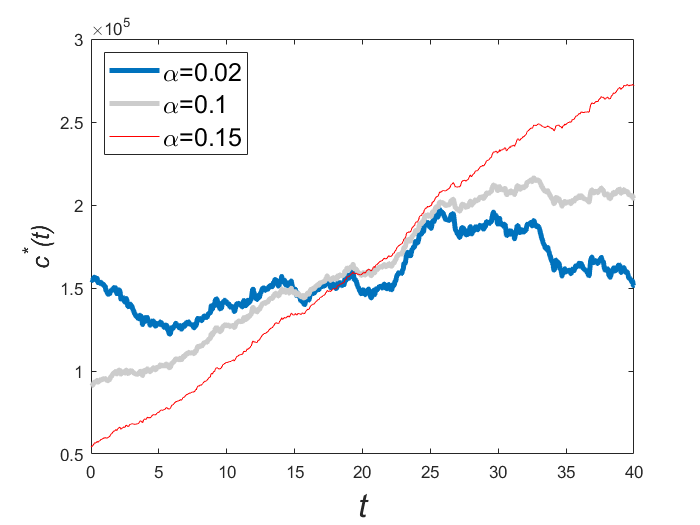}\\
			\vspace{0.02cm}
			
		\end{minipage}%
	}%
	\subfigure[]{
		\begin{minipage}[t]{0.32\textwidth}
			\centering
			\includegraphics[width=\textwidth]{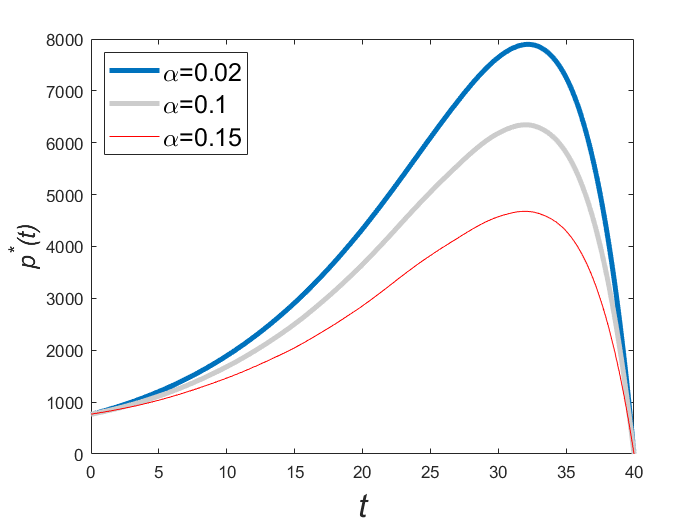}\\
			\vspace{0.02cm}
			
		\end{minipage}%
	}%
	\centering
	\caption{ The influence of consumption habits parameter $\alpha$ on household investment strategy $\pi^*$, consumption strategy $c^*$, and life insurance strategy $p^*$. }
	\vspace{-0.2cm}
\end{figure}

According to equation (2.1), the consumption habits model has nothing to do with the health status of the breadwinner.
In this section, the influence of consumption habits on the optimal strategies is discussed by taking the health state of the breadwinner as an example.
Figure 5.5 shows the influence of consumption habits parameter $\alpha$ on household optimal investment, consumption, and life insurance decisions.
As can be seen from Figure 5.5 (a), as parameter $\alpha$ increases, the amount of household wealth invested in risky assets decreases.
Parameter $\alpha$ measures the influence of historical consumption levels on current consumption behavior.
When $\alpha$ is large, that is, households attach more importance to maintaining current consumption levels.
Therefore, households will invest more prudently and invest less in risky assets.
As can be seen from Figure 5.5 (b), the larger parameter $\alpha$, the lower the initial household consumption and the faster the growth of consumption expenditure.
This is because when $\alpha$ is large, current consumption behavior is more influenced by past consumption levels.
Therefore, in order to avoid future consumption that is too high and exceeds expectations, households reduce consumption at an early stage to ensure that future consumption habits are met.
As can be seen from Figure 5.5 (c), the amount of life insurance decreases with the increase of parameter $\alpha$.
This is because under a larger $\alpha$, the breadwinner must secure sufficient financial wealth to sustain the household until his/her death, thus reducing the incentive to bequeath, and  the purchase of life insurance.

\begin{figure}
	\centering
	\subfigure[]{
		\begin{minipage}[t]{0.32\textwidth}
			\centering
			\includegraphics[width=\textwidth]{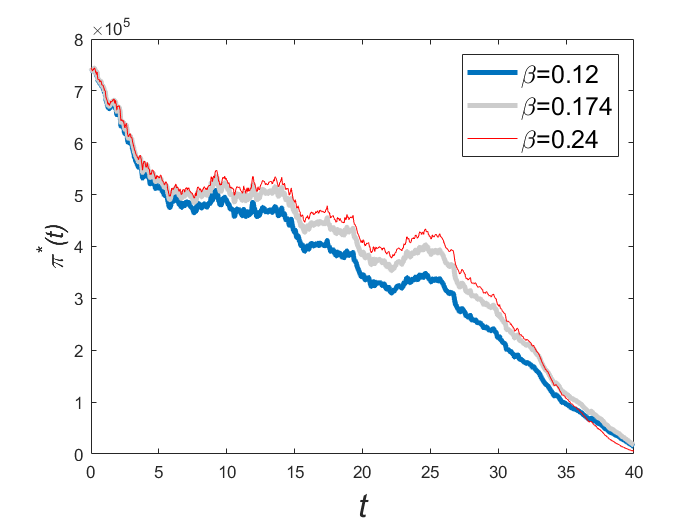}\\
			\vspace{0.02cm}
			
		\end{minipage}%
	}%
	\subfigure[]{
		\begin{minipage}[t]{0.32\textwidth}
			\centering
			\includegraphics[width=\textwidth]{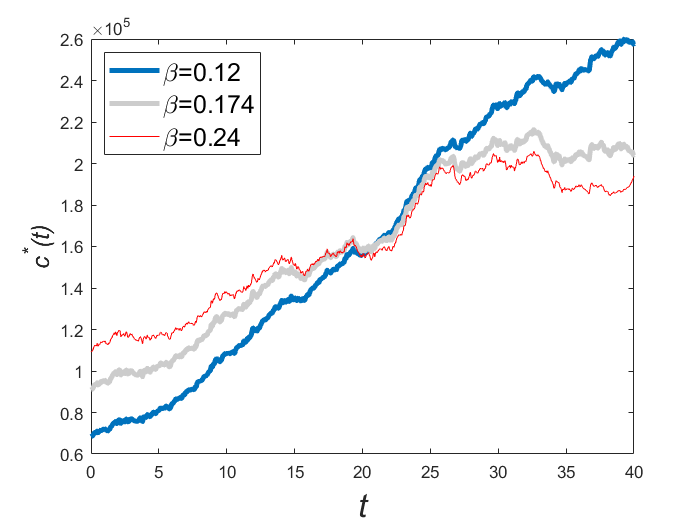}\\
			\vspace{0.02cm}
			
		\end{minipage}%
	}%
	\subfigure[]{
		\begin{minipage}[t]{0.32\textwidth}
			\centering
			\includegraphics[width=\textwidth]{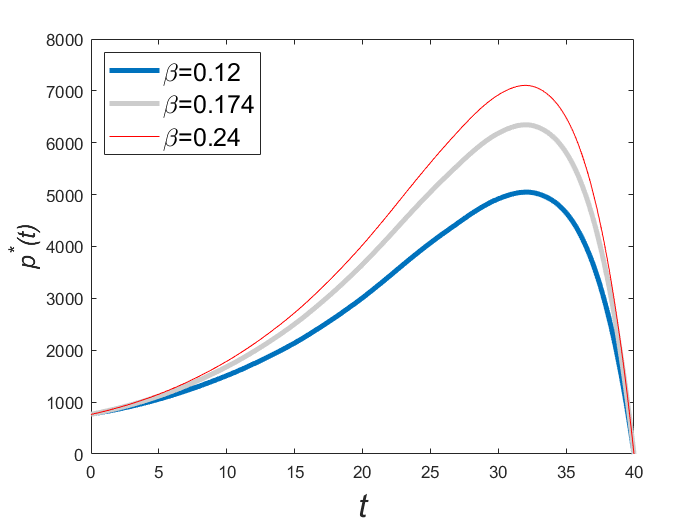}\\
			\vspace{0.02cm}
			
		\end{minipage}%
	}%
	\centering
	\caption{ The influence of consumption habits parameter $\beta$ on household investment strategy $\pi^*$, consumption strategy $c^*$, and life insurance strategy $p^*$.}
	\vspace{-0.2cm}
\end{figure}

Figure 5.6 shows the influence of consumption habits parameter $\beta$ on household optimal strategies.
In contrast to the effect of parameter $\alpha$ shown in Figure 5.5, the consumption habits parameter $\beta$ measures the extent to which households forget past consumption.
When the $\beta$ value is large, when the household forms the current consumption habits, the past consumption level has less impact on the current consumption behavior, and family members pay more attention to the current quality of life and consumption level.
Therefore, with the increase of parameter $\beta$, households are more inclined to increase investment and early consumption expenditure, and will increase life insurance to ensure that the expected consumption needs of dependents can be met in the future, and provide additional protection for household heritage.

With reference to Kraft et al.(2017), the value of $\beta-\alpha$ can be used to measure the strength of consumption habits, and the smaller the value of $\beta-\alpha$, the stronger the consumption habits.
By observing Figures 5.5 and 5.6, it can be concluded that households with stronger consumption habits have more conservative investment strategies, faster growth of consumption expenditure and more life insurance.

\section{Concluding}

For households consisting of one breadwinner and one dependant, this paper considers the health status of the breadwinners in the framework of investment, consumption and life insurance issues. Health shocks affect financial decisions by reducing households' income and increasing the death rate of the breadwinners.
In addition, considering the consumption habits of households, the differences of households asset allocation strategies under different consumption habits are compared.
According to the principle of dynamic programming, the corresponding HJB equation is established and solved, and the optimal household investment, consumption and life insurance strategies are obtained.
Finally, an example of critical illness model is given to demonstrate the application of the research content in this paper.
Through the sensitivity analysis of the important parameters, the following findings are obtained.
(\rmnum{1}) Households with breadwinners in poor health may adopt conservative investment strategies due to increased medical expenditures, reducing consumption expenditure to better manage households' wealth. In order to protect dependants, households suffering health shocks may need to take out more life insurance.
(\rmnum{2}) Critical illness of breadwinners may lead to a reduction in household income, which directly affects the households' asset allocation strategies. Households with less labor income spend less on investment, consumption, and life insurance.
(\rmnum{3}) Households with stronger consumption habits are more susceptible to past consumption levels.
A conservative investment strategy and lower early consumption can be adopted to cope with higher consumption in the future.
Higher levels of consumption habits will also reduce the demand for life insurance.

\vspace{0.2cm}

Nonetheless, there are some limitations to this study. For example, the assumption is made that the income of the breadwinners in a healthy state follows an exponential growth pattern. Subsequent research efforts may delve into exploring alternative models that are more closely aligned with the actual dynamics of income.

\vspace{0.5cm}

\noindent{\bf \Large  Declaration of competing interest}

\vspace{0.3cm}

There is no competing interest.

\vspace{0.3cm}

\noindent{\bf \Large Acknowledgements}

\vspace{0.3cm}

This work is supported by the National Natural Science Foundation of China (Nos: 11961064).

\section*{Appendix}
\setcounter{equation}{0}
\setcounter{subsection}{0}
\renewcommand{\theequation}{A.\arabic{equation}}
\renewcommand{\thesubsection}{A.\arabic{subsection}}

In this appendix, we provide all the proofs of main results, as well as lemmas, of this paper.

\vspace{0.2cm}

\begin{proof}[\bf Proof of Theorem 3.2]
For the HJB equation
\begin{equation}
		\begin{aligned}
			\mathop{{\rm sup}}\limits_{{\nu_d}\in \Pi_d}&\Big\{V_{d,t}+V_{d,x}[rx_d+\pi_d(\mu-r)-c_d]+V_{d,h}(\alpha c_d-\beta h_d) +\frac{1}{2} \sigma^2\pi_d^2 V_{d,xx} \\
			& +k_d {\rm e}^{-\rho t}\frac{(c_d-h_d)^{1-\gamma}}{1-\gamma} \Big\}=0.
		\end{aligned}
	\end{equation}
By first order condition
\begin{equation}
		\left\{ {\begin{array}{*{20}{l}}
				{\pi_d^*(t)=\displaystyle{-\frac{(\mu-r)V_{d,x}}{\sigma^2V_{d,xx}}},}\\
				{c_d^*(t)=\displaystyle{h_d+(k_d {\rm e}^{-\rho t})^{\frac{1}{\gamma}}(V_{d,x}-\alpha V_{d,h})^{-\frac{1}{\gamma}},}}
		\end{array}} \right.
	\end{equation}
The partial differential equation about $V_d(t,x_d,h_d)$ is obtained by substituting the optimal solution $(A.2)$ into HJB equation $(A.1)$
\begin{equation}\label{eqn-A.3}
		\begin{aligned}
			V_{d,t}-\frac{(\mu-r)^2V_{d,x}^2}{2\sigma^2V_{d,xx}}+(rx_d-h_d)V_{d,x}+(\alpha - \beta)h_dV_{d,h}+
			\frac{\gamma}{1-\gamma}(k_d {\rm e}^{-\rho t})^{\frac{1}{\gamma}}(V_{d,x}-\alpha V_{d,h})^{1-\frac{1}{\gamma}} =0.
		\end{aligned}
	\end{equation}

Suppose the solution of the equation has the following form
\begin{equation}
		\displaystyle{V_d(t,x_d,h_d)=\frac{1}{1-\gamma}[g(t)]^{\gamma}[x_d-h_dB(t)]^{1-\gamma}.}
	\end{equation}
The corresponding partial derivative is
\begin{equation}
		\begin{aligned}
			& V_{d,t} = -h_d B^{'}(t)(x_d-h_d B)^{-\gamma}g^{\gamma}+\frac{\gamma}{1-\gamma}(x_d-h_d B)^{1-\gamma}g^{\gamma-1}g^{'}(t) ,\\
			& V_{d,x} = (x_d-h_d B)^{-\gamma}g^{\gamma},\\
			& V_{d,h} = -B(x_d-h_dB)^{-\gamma}g^{\gamma},\\
			& V_{d,xx} = -\gamma(x_d-h_d B)^{-\gamma-1}g^{\gamma}.
		\end{aligned}	
	\end{equation}
By substituting $(A.4)$ and $(A.5)$ into $(A.3)$ and simplifying, we get
\begin{equation}
		\begin{aligned}
			&h_d(x_d-h_d B)^{-\gamma}g^{\gamma}[-B^{'}(t)+(r+\beta-\alpha)B-1] \\
			&+\frac{\gamma}{1-\gamma}g^{\gamma-1}(x_d-h_d B)^{1-\gamma}\Big[g^{'}(t)
			+\frac{1-\gamma}{\gamma}[r+\frac{(\mu-r)^2}{2\sigma^2\gamma}]g(t)+(k_d {\rm e}^{-\rho t})^{\frac{1}{\gamma}}(1+\alpha B)^{1-\frac{1}{\gamma}}\Big]=0.
			\nonumber
		\end{aligned}
	\end{equation}
Obtain a system of differential equations
\begin{equation}
		\left\{ {\begin{array}{*{20}{l}}
				{\displaystyle{B^{'}(t)=(r+\beta-\alpha)B-1,}}\\
				{\displaystyle{g^{'}(t)=-\frac{1-\gamma}{\gamma}[r+\frac{(\mu-r)^2}{2\sigma^2\gamma}]g(t)-(k_d {\rm e}^{-\rho t})^{\frac{1}{\gamma}}(1+\alpha B)^{1-\frac{1}{\gamma}},}}
		\end{array}} \right.
		\nonumber
	\end{equation}
Combined with boundary condition $B(T)=0$ and $g(T)=(\omega_d {\rm e}^{-\rho T})^{\frac{1}{\gamma}}$, equation $(3.8)$ can be obtained, substitute $(A.5)$ into $(A.2)$ to get equation $(3.7)$.
\end{proof}

\begin{proof}[\bf Proof of Lemma 3.1]
The objective function (3.9) can be written as
\begin{equation}
		\begin{aligned}	
			&J(t,x_a,h_a,i;\nu_a(\cdot)) \\
			=&E_{t,x_a,h_a,i}\Big[\int_{t}^{\tau_i \land T}U_a(s,c_a(s),h_a(s),\eta(s)){\rm d}s+\mathbb{I}_{\{\tau_i > T \}}\Psi_a(T,X_a(T),\eta(T))\\
			&+\mathbb{I}_{\{\tau_i < T \}}[\int_{\tau_i}^{T}U_d(s,c_d(s),h_d(s)){\rm d}s+\Psi_d(T,X_d(T))]\Big]\\
			=&E_{t,x_a,h_a,i}\Big[\int_{t}^{\tau_i \land T}U_a(s,c_a(s),h_a(s),\eta(s)){\rm d}s+\mathbb{I}_{\{\tau_i > T \}}\Psi_a(T,X_a(T),\eta(T))\\
			&+\mathbb{I}_{\{\tau_i < T \}}V_d(\tau_i,x_a(\tau_i)+\frac{p_a(\tau_i,i)}{\lambda(\tau_i,i)},h_d(\tau_i))\Big].
			\nonumber
		\end{aligned}	
	\end{equation}	
Where
\begin{equation}
		\begin{aligned}
			&E_{t,x_a,h_a,i}\Big[\int_{t}^{\tau_i \land T}U_a(s,c_a(s),h_a(s),\eta(s)){\rm d}s\Big]\\
			=&E_{t,x_a,h_a,i}\Big[\mathbb{I}_{\{\tau_i < T\}}\int_{t}^{\tau_i}U_a(s,c_a,h_a,\eta(s)){\rm d}s+\mathbb{I}_{\{\tau_i > T\}}\int_{t}^{T}U_a(s,c_a,h_a,\eta(s)){\rm d}s\Big]\\
			=&E_{t,x_a,h_a,i}\Big\{\int_{t}^{T}f(u,t,\eta(u)){\rm d}u \int_{t}^{u}U_a(s,c_a,h_a,\eta(s)){\rm d}s\\
            &+\Big[1-\int_{t}^{T}f(u,t,\eta(u)){\rm d}u\Big]\int_{t}^{T}U_a(s,c_a,h_a,\eta(s)){\rm d}s\Big\}\\
			=&E_{t,x_a,h_a,i}\Big[\bar{F}(s,t,i)\int_{t}^{T}U_a(s,c_a,h_a,\eta(s)){\rm d}s\Big].	
			\nonumber
		\end{aligned}	
	\end{equation}
In a similar way
\begin{equation}
		\begin{aligned}
			E_{t,x_a,h_a,i}\Big[\mathbb{I}_{\{\tau_i > T\}}\Psi_a(T,X_a(T),\eta(T))\Big]
=E_{t,x_a,h_a,i}\Big[\bar{F}(T,t,i)\Psi_a(T,X_a(T),\eta(T))\Big],
			\nonumber
		\end{aligned}	
	\end{equation}

\begin{equation}
		\begin{aligned}
			&E_{t,x_a,h_a,i}\Big[\mathbb{I}_{\{\tau_i < T\}}V_d\Big(\tau_i,x_a(\tau_i)+\frac{p_a(\tau_i,i)}{\lambda(\tau_i,i)},h_d(\tau_i)\Big)\Big]\\
=&E_{t,x_a,h_a,i}\Big[\int_{t}^{T}f(s,t,i)V_d\Big(s,x_a(s)+\frac{p_a(s,\eta(s))}{\lambda(s,\eta(s))},h_d(s)\Big){\rm d}s\Big].
			\nonumber
		\end{aligned}	
	\end{equation}
Where $\bar{F}(s,t,i)$ and $f(s,t,i)$ denote the conditional survival function and the conditional probability density function, respectively, for a breadwinner with the physical state of $i$, satisfied
$$\displaystyle{\bar{F}(s,t,i)=\frac{\bar{F}(s,i)}{\bar{F}(t,i)},\qquad f(s,t,i)=\frac{f(s,i)}{\bar{F}(t,i)}}.$$
Therefore, equation (3.12) can be obtained.
\end{proof}

\begin{proof}[\bf Proof of Theorem 3.4]
For the HJB equation
\begin{equation}\label{eqn-C.1}
		\begin{aligned}
			\mathop{{\rm sup}}\limits_{{\nu_a}\in \Pi_a}&\Big\{\tilde{V}_{t}+\tilde{V}_{x}[rx_a+\pi_a(\mu-r)+y(t,i)-p_a(t,i)-c_a]+\tilde{V}_{h}(\alpha c_a-\beta h_a) +\frac{1}{2} \sigma^2\pi_a^2 \tilde{V}_{xx} \\
			& +k_{a,i}\bar{F}(t,i) {\rm e}^{-\rho t}\frac{(c_a-h_a)^{1-\gamma}}{1-\gamma} +\frac{1}{1-\gamma}f(t,i)[g(t)]^{\gamma}\left[x_a+\frac{p_a(t,i)}{\lambda(t,i)}-h_d B(t)\right]^{1-\gamma}\\
			&+\sum\limits_{j\in \mathbb{S};j\ne i} {q_{ij}(t)[\tilde{V}(t,x_a,h_a,j)-\tilde{V}(t,x_a,h_a,i)]} \Big\}=0.
		\end{aligned}
	\end{equation}
By first order condition
\begin{equation}\label{eqn-C.2}
		\left\{ {\begin{array}{*{20}{l}}
				{\pi_a^*(t)=\displaystyle{-\frac{(\mu-r)\tilde{V}_{x}}{\sigma^2\tilde{V}_{xx}}},}\\
				{c_a^*(t)=\displaystyle{h_a+[k_{a,i} \bar{F}(t,i) {\rm e}^{-\rho t}]^{\frac{1}{\gamma}}(\tilde{V}_{x}-\alpha \tilde{V}_{h})^{-\frac{1}{\gamma}},}}\\
				{p_a^*(t)=\displaystyle{\lambda(t,i)\Big[h_dB(t)-x_a
+[\lambda(t,i)\tilde{V}_x]^{-\frac{1}{\gamma}}[f(t,i)]^{\frac{1}{\gamma}}g(t)\Big]}.}
		\end{array}} \right.
	\end{equation}
The partial differential equation about $\tilde{V}(t,x_a,h_a,i)$ is obtained by substituting the optimal solution $(A.7)$ into HJB equation $(A.6)$
\begin{equation}\label{eqn-C.3}
		\begin{aligned}
			&\tilde{V}_{t}-\frac{(\mu-r)^2\tilde{V}_{x}^2}{2\sigma^2\tilde{V}_{xx}}+[rx_a-h_a+y(t,i)-\lambda(t,i) h_d B+\lambda(t,i) x_a]\tilde{V}_{x}+(\alpha - \beta)h_a\tilde{V}_{h}\\
			&+\frac{\gamma}{1-\gamma}[k_{a,i}\bar{F}(t,i) {\rm e}^{-\rho t}]^{\frac{1}{\gamma}}(\tilde{V}_{x}-\alpha \tilde{V}_{h})^{1-\frac{1}{\gamma}}+\frac{\gamma}{1-\gamma}[f(t,i)]^{\frac{1}{\gamma}}g(t)[\lambda(t,i) \tilde{V}_x]^{1-\frac{1}{\gamma}}\\
			&+\sum\limits_{j\in \mathbb{S};j\ne i} {q_{ij}(t)[\tilde{V}(t,x_a,h_a,j)-\tilde{V}(t,x_a,h_a,i)]} =0.
		\end{aligned}
	\end{equation}

Suppose the solution of the equation has the following form
\begin{equation}\label{eqn-C.4}
		\displaystyle{\tilde{V}(t,x_a,h_a,i)=\frac{1}{1-\gamma}[G_i(t)]^{\gamma}[x_a+M(t)+h_aA(t)]^{1-\gamma}.}
	\end{equation}
The corresponding partial derivative is
\begin{equation}\label{eqn-C.5}
		\begin{aligned}
			& \tilde{V}_{t} = [M^{'}(t)+h_a A^{'}(t)](x_a+M+h_a A)^{-\gamma}[G_i(t)]^{\gamma}+\frac{\gamma}{1-\gamma}(x_a+M+h_a A)^{1-\gamma}[G_i(t)]^{\gamma-1}G_i^{'}(t) ,\\
			& \tilde{V}_{x} = (x_a+M+h_a A)^{-\gamma}[G_i(t)]^{\gamma},\\
			& \tilde{V}_{h} = A(x_a+M+h_aA)^{-\gamma}[G_i(t)]^{\gamma},\\
			& \tilde{V}_{xx} = -\gamma(x_a+M+h_a A)^{-\gamma-1}[G_i(t)]^{\gamma}.
		\end{aligned}	
	\end{equation}
By substituting $(A.9)$ and $(A.10)$ into $(A.8)$ and simplifying, we get
\begin{equation}
		\begin{aligned}
			&h_a(x_a+M+h_a A)^{-\gamma}[G_i(t)]^{\gamma}\Big[A^{'}(t)-[r+\lambda(t,i)]A-1+(\alpha-\beta)A\Big] \\
			&+(x_a+M+h_a A)^{-\gamma}[G_i(t)]^{\gamma}\Big[M^{'}(t)+y(t,i)-[r+\lambda(t,i)]M-\lambda(t,i)h_dB\Big] \\
			&+\frac{\gamma}{1-\gamma}[G_i(t)]^{\gamma-1}(x_a+M+h_a A)^{1-\gamma}\Big[G_i^{'}(t)
			+\frac{1}{\gamma}\Big((1-\gamma)[r+\lambda(t,i)+\frac{(\mu-r)^2}{2\sigma^2\gamma}]\\
			&+\sum\limits_{j\in \mathbb{S};j\ne i} {q_{ij}(t)[(\frac{G_j(t)}{G_i(t)})^{\gamma}-1]} \Big)G_i(t)+[k_{a,i}\bar{F}(t,i){\rm e}^{-\rho t}]^{\frac{1}{\gamma}}(1-\alpha A)^{1-\frac{1}{\gamma}}\\
&+[f(t,i)]^{\frac{1}{\gamma}}[\lambda(t,i)]^{1-\frac{1}{\gamma}}g(t)\Big]=0.
			\nonumber
		\end{aligned}
	\end{equation}
The systems of differential equations are obtained
\begin{equation}
		\begin{aligned}
			&\left\{ {\begin{array}{*{20}{l}}
					{\displaystyle{A^{'}(t)=[r+\beta-\alpha+\lambda(t,i)]A+1,}}\\
					{A(T)=0.}
			\end{array}} \right. \\
			&\left\{ {\begin{array}{*{20}{l}}
					{\displaystyle{M^{'}(t)=[r+\lambda(t,i)]M+\lambda(t,i)h_dB-y(t,i),}}\\
					{M(T)=0.}
			\end{array}} \right. \\
			&\left\{ {\begin{array}{*{20}{l}} {G_i^{'}(t)+\frac{1}{\gamma}\Big[(1-\gamma)[r+\lambda(t,i)+\frac{(\mu-r)^2}{2\sigma^2\gamma}]+\sum\limits_{j\in \mathbb{S};j\ne i} {q_{ij}(t)[(\frac{G_j(t)}{G_i(t)})^{\gamma}-1]} \Big]G_i(t)} \\
				{\quad +[k_{a,i}\bar{F}(t,i){\rm e}^{-\rho t}]^{\frac{1}{\gamma}}(1-\alpha A)^{1-\frac{1}{\gamma}}+[f(t,i)]^{\frac{1}{\gamma}}[\lambda(t,i)]^{1-\frac{1}{\gamma}}g(t)=0,}\\
				{G_i(T)=[\omega_{a,i}\bar{F}(T,i){\rm e}^{-\rho T}]^{\frac{1}{\gamma}}.}
		\end{array}} \right.
        \nonumber
		\end{aligned}
	\end{equation}
whose solution is
\begin{equation}
		\begin{aligned}
			A(t) &= \displaystyle{-\int_{t}^{T}{\rm e}^{-\int_{t}^{s}[r+\lambda(u,\eta(u))+\beta-\alpha]{\rm d}u}{\rm d}s,} \\
			M(t) &= \displaystyle{\int_{t}^{T}{\rm e}^{-\int_{t}^{s}[r+\lambda(u,\eta(u))]{\rm d}u}[y(s,\eta(s))-\lambda(s,\eta(s))h_dB(s)]{\rm d}s.}
		\end{aligned}
		\nonumber	
	\end{equation}

\end{proof}

\vspace{0.3cm}


\vspace{0.2cm}

\newpage

\begin{thebibliography}{111}
	
	\bibitem{1} Andersen, S., Harrison, G.W., Lau, M.I., Rutstrm, E., 2008. Eliciting risk and time preferences. Econometrica 76, 583-618. doi:10.1111/j.14680262.2008.00848.x.
	
	\bibitem{2} Barucci, E., Biffis, E., Marazzina, D., 2023. Health insurance, portfolio
 choice, and retirement incentives. European Journal of Operational Research 307,
 910-921. doi:10.1016/j.ejor.2022.09.016.
	
	\bibitem{3} Bayraktar, E., Young, V.R., 2013. Life Insurance Purchasing to Maximize Utility of Household Consumption. North American Actuarial Journal 17(2), 1-22. doi:10.1080/10920277.2013.793159.
	
	\bibitem{4} Ben-Arab, M., Briys, E., Schlesinger, H., 1996. Habit formation and the demand for insurance.
	The Journal of Risk and Insurance 63(1), 111-119. doi:10.2307/253519.
	
	\bibitem{5} Berkowitz, M., Qiu, J., 2006. A further look at household portfolio choice and health status.
	Journal of Banking and Finance 30, 1201-1217. doi:10.1016/j.jbank n.2005.05.006.
	
	\bibitem{6} Boyle, P., Tan, K.S., Wei, P., Zhuang, S.C., 2022. Annuity and insurance choice under habit formation. Insurance: Mathematics and Economics 105, 211-237. doi:10.1016/j.in smatheco.2022.04.003.
	
	\bibitem{7} Bruhn, K., Steffensen, M., 2011. Household consumption, investment and life insurance.
Insurance: Mathematics and Economics 48(3), 315-325. doi:10.1016/j.in smatheco.2010.12.004.
	
	\bibitem{8} Charupat, N., Milevsky, M.A., 2002. Optimal asset allocation in life annuities: a note.
Insurance: Mathematics and Economics 30(2), 199-209. doi:10.1016/s0167-6687(02)0 0097-5.
	
	\bibitem{9} Decker, S., Schmitz, H., 2016. Health shocks and risk aversion. Journal of Health Economics 50, 156-170. doi:10.1016/j.jhealeco.2016.09.006.
	
	\bibitem{10} Detemple, J.B., Zapatero, F., 1992. Optimal consumption-portfolio policies with habit formation.
	Mathematical Finance 2(4), 251-274. doi:10.1111/j.1467-9965.1992.tb00 032.x.
	
	\bibitem{11} Duesenberry, J.S., 1949. Income Saving and the Theory. Harvard University Press.
	
	\bibitem{12} Edwards, R.D., 2008. Health risk and portfolio choice.
Health Economics 26(4), 472-485. doi:10.1198/073500107000000287.
	
	\bibitem{13} Halla, M., Zweimller, M., 2013. The effect of health on earnings: quasi-experimental evidence from commuting accidents. Labour Economics 24, 23-38.
	
	\bibitem{14} Hambel, C., Kraft, H., Schendel, L.S., Steffensen, M., 2017. Life insurance demand under health shock risk. Journal of Risk and Insurance 84(4), 1171-1202. doi:10.1111/j ori.12149.
	
	\bibitem{15} Hambel, C., 2020. Health shock risk, critical illness insurance, and housing services.
	Insurance: Mathematics and Economics 91, 111-128. doi:10.1016/j.insmatheco.2 020.01.008.
	
	
	\bibitem{16} Hicks, J., 1965. Capital and Growth. Oxford University Press.
	
	\bibitem{17} Kraft, H., Munk, C., Seifried, F.T., Wagner, S., 2017. Consumption habits and humps. Economic Theory 64(2), 305-330. doi:10.1007/s00199-016-0984-1.
	
	\bibitem{18} Kraft, H., Weiss, F., 2023. Pandemic portfolio choice. European Journal of
Operational Research 305, 451-462. doi:10.1016/j.ejor.2022.05.035.
	
	\bibitem{19} Kwak, M., Yong, H.S., Choia, U.J., 2011. Optimal investment and consumption decision of a family with life insurance. Insurance: Mathematics and Economics 48(2), 176-188. doi:10.1016/j.insmatheco.2010.10.012.
	
	\bibitem{20} Lee, H., Cha, J.H., 2018. A dynamic bivariate common shock model with cumulative effect and its actuarial application. Scandinavian Actuarial Journal 10, 890-906. doi:10.1080/03461238.2018.1470562.
	
	\bibitem{21} Lenhart, O., 2019. The effects of health shocks on labor market outcomes:evidence from UK panel data. European Journal of Health Economics 20, 83-98. doi:10.1007/s10198-018-0985-z.
	
	\bibitem{22} Liang, Z.X., Zhao, X.Y., 2016. Optimal mean-variance efficiency of a family with life insurance under inflation risk. Insurance: Mathematics and Economics 71, 164-178. doi:10.1016/j.insmatheco.2016.09.004.
	
	\bibitem{23} Liu, J.Z., Lin, L.Y., Meng, H., 2020. Optimal consumption, life insurance and investment decision with habit formation. Acta Mathematicae Applicatae Sinica 43(03), 517-534. doi:10.12387/C2020040.
	
	\bibitem{24} Love, D.A., 2009. The effects of marital status and children on savings and portfolio choice. The Review of Financial Studies 23(1), 385-432. doi:10.1093/rfs/hhp020.
	
	\bibitem{25} Luciano, E., Spreeuw, J., Vigna, E., 2008. Modelling stochastic mortality for dependent lives. Insurance: Mathematics and Economics 43(2), 234-244.  doi:10.1016/j.in smatheco.2008.06.005.
	
	\bibitem{26} Moore, K.S., Young, V.R., 2006. Optimal insurance in a continuous-time model.
Insurance: Mathematics and Economics 39(1), 47-68. doi:10.1016/j.insmatheco.2006.01.009.
	
	\bibitem{27} Mousa, A.S., Pinheiro, D., Pinto, A.A., 2016. Optimal life-insurance selection and purchase within a market of several life-insurance providers. Insurance: Mathematics and Economics 67, 133-141. doi:10.1016/j.insmatheco.2016.01.002.
	
	\bibitem{28} Nielsen, P.H., Steffensen, M., 2008. Optimal investment and life insurance strategies under minimum and	maximum constraints. Insurance: Mathematics and Economics 43(1),
15-28. doi:10.1016/j.insmatheco.2007.09.007.
	
	\bibitem{29} Peng, X.C., Li, B.H., 2023. Optimal investment, consumption and life insurance purchase with learning about return predictability. Insurance: Mathematics and Economics 113,
70-95. doi:10.1016/j.insmatheco.2023.07.005.
	
	\bibitem{30} Pirvu, T.A., Zhang, H., 2012. Optimal investment, consumption and life insurance under mean-reverting returns: the complete market solution. Insurance: Mathematics and Economics 51,
303-309. doi:10.1016/j.insmatheco.2012.05.002.
	
	\bibitem{31} Polkovnichenko, V., 2006. Life-cycle portfolio choice with additive habit formation preferences and uninsurable labor income risk. The Review of Financial Studies 20(1), 83-124. doi:10.1093/rfs/hhl006.
	
	\bibitem{32} Pollak, R.A., 1970. Habit Formation and Dynamic Demand Functions. Journal of Political
Economy 78, 745-763. doi:10.1086/259667.
	
	\bibitem{33} Rice, N., Robone, S., 2022. The effects of health shocks on risk preferences: Do personality traits matter? Journal of Economic Behavior and Organization 204, 356-371. doi:10.1016/j.jebo.2022.10.016.
	
	\bibitem{34} Richard, S.F., 1975. Optimal consumption, portfolio and life insurance rules for an uncertain lived individual in a continuous time model. Journal of Financial Economics 2,
187-203. doi:10.1016/0304-405x(75)90004-5.
	
	\bibitem{35} Riphahn, R.T., 1999. Income and employment effects of health shocks: a test case for the German welfare state. Journal of Popolation Economics 12(3), 363-389. doi:10.1007/s001480050104.
	
	\bibitem{36} Ryder, H.E., Heal, G.M., 1973. Optimal Growth with Intertemporally Dependent Preferences. Review of Economic Studies 40, 1-33. doi:10.2307/2296736.
	
	\bibitem{37} Shen, Y., Sherris, M., 2018. Lifetime asset allocation with idiosyncratic and systematic mortality risks. Scand Actuar 4, 294-327. doi:10.1080/03461238.2017.1343749.
	
	\bibitem{38} Sundaresan, S.M., 1989. Intertemporally dependent preferences and the volatility of consumption and wealth.	The Review of Financial Studies 2(1), 73-89. doi:10.1093/rfs/2.1.73.
	
	\bibitem{39} Tao, C., Rong, X.M., Zhao, H., 2023. Stochastic control with inhomogeneous regime
switching: Application	to consumption and investment with unemployment and	reemployment.
Journal of Mathematical Economics 107, 102849. doi:10.10 16/j.jmateco.2023.102849.
	
	\bibitem{40} Wang, N., Jin, Z., Siu, T.K., Qiu, M., 2021. Household consumptioninvestment-insurance decisions with uncertain income and market ambiguity. Scandinavian Actuarial Journal 10,
832-865. doi:10.1080/03461238.2021.1886981.
	
	\bibitem{41} Wang, Y.J., Jin, Z.J., Yuan, Y., 2023. The consequences of health shocks on households: Evidence from China. China Economic Review 79, 101969. doi:10.10 16/j.chieco.2023.101969.
	
	\bibitem{42} Wei, J.Q., Cheng, X., Jin, Z., Wang, H., 2020. Optimal consumption-investment and life-insurance purchase strategy for couples with correlated lifetimes. Insurance: Mathematics and Economics 91,
244-256. doi:10.1016/j.insmatheco.2020.02.006.
	
	\bibitem{43} Yaari, M.E., 1965. Uncertain lifetime, life insurance, and the theory of the
consumer. The Review of Economic Studies 32(2), 137-150. doi:10.2307/2296058.
	
	\bibitem{44} Ye, J.C., 2006. Optimal Life Insurance Purchase, Consumption, and Portfolio under an Uncertain Life. Ph.D. University of Illinois at Chicago.
	
	\bibitem{45} Ye, J.C., 2019. Stochastic utilities with subsistence and satiation: Optimal life insurance purchase, consumption and investment. Insurance: Mathematics and Economics 89,
193-212. doi:10.1016/j.insmatheco.2019.10.008.
	
\end{thebibliography}
\end{document}